\documentclass[10pt,twocolumn,letterpaper]{article}
\usepackage[T1]{fontenc}
\usepackage{times}
\usepackage[letterpaper,textwidth=6.75in,textheight=9in,centering,columnsep=0.25in]{geometry}
\usepackage{microtype}
\usepackage{graphicx}
\usepackage{amsmath,amssymb}
\usepackage{booktabs,tabularx}
\usepackage{multirow}
\usepackage{enumitem}
\usepackage{xspace}
\usepackage{xcolor}
\definecolor{grey}{rgb}{0.5,0.5,0.5}
\definecolor{pink}{rgb}{0.8,0.2,0.6}
\usepackage{pifont}
\usepackage{listings}
\usepackage{url}
\usepackage[authoryear,round]{natbib}
\usepackage{algorithm}
\usepackage{algorithmic}
\usepackage[font=small,labelfont=bf]{caption}
\usepackage[colorlinks=true,allcolors=blue]{hyperref}
\hypersetup{
  pdftitle={Speculative Ad-hoc Querying},
  pdfauthor={Haoyu Li; Srikanth Kandula; Maria Angels de Luis Balaguer; Aditya Akella; Venkat Arun},
  pdfsubject={Analytical SQL and speculative execution},
  pdfkeywords={Agentic Systems, Speculative Execution, SQL, Large Language Models}
}
\let\cite\citep

\newcommand{\sys}{SpeQL\xspace}

\newcommand{\sql}[1]{\textcolor{blue}{\texttt{#1}}}
\newcommand{\para}[1]{\smallskip\noindent\textbf{\textit{#1}}}
\title{Speculative Ad-hoc Querying}
\author{\begin{tabular}{ccc}
\normalsize Haoyu Li & \normalsize Srikanth Kandula & \normalsize Maria Angels de Luis Balaguer \\
\small The University of Texas at Austin & \small Amazon Web Services & \small Microsoft Research \\
\small \href{mailto:lhy@utexas.edu}{\texttt{lhy@utexas.edu}} &
\small \href{mailto:kandula@gmail.com}{\texttt{kandula@gmail.com}} &
\small \href{mailto:angelsd@microsoft.com}{\texttt{angelsd@microsoft.com}}
\end{tabular}\\[1em]
\begin{tabular}{cc}
\normalsize Aditya Akella & \normalsize Venkat Arun \\
\small The University of Texas at Austin & \small The University of Texas at Austin \\
\small \href{mailto:akella@cs.utexas.edu}{\texttt{akella@cs.utexas.edu}} &
\small \href{mailto:venkat@utexas.edu}{\texttt{venkat@utexas.edu}}
\end{tabular}}
\date{}

\begin{document}
\twocolumn[
\begin{@twocolumnfalse}
\maketitle
\begin{abstract}

Large language models (LLMs) invite substantial effort and investment, yet agentic systems can spend much of their
time waiting for tools. This paper explores how a small auxiliary
LLM can reduce these waits by speculatively preparing a database
while the primary LLM reasons.

The challenge is that the auxiliary model must anticipate the
primary model's actions before the primary model takes them.
Our key insight is that database systems give the speculator a
structural advantage: it need not predict the exact next query.
Instead, it can prepare work shared by many plausible queries.
Predicting query shapes enables advance compilation, predicting
data access warms caches, and predicting broader data requirements
enables precomputation of superset results that subsequent queries
can refine.

We present SpeQL, a system that exploits this advantage for
analytical SQL workloads. SpeQL constructs reusable superset
queries, materializes their components, and rewrites subsequent
queries against the resulting temporary tables. Its scheduler
prioritizes foreground queries while using reasoning intervals
for speculative preparation. The same engine supports humans
writing SQL, using incomplete queries to guide speculation and
displaying intermediate results as they type.

In a replay-based evaluation of agent workflows, SpeQL reduces
SQL execution time by up to 31.8\% and combined primary-agent
reasoning and active SQL time by up to 22.6\%, with approximately
6\% higher recorded main-agent and prediction API cost and
10\% more read I/O on the full dataset. A separate 24-participant
study finds significantly lower query latency on both analytical
tasks and improved time-to-answer on one, with 87.5\% of
participants reporting increased productivity.

\end{abstract}
\vspace{1em}
\end{@twocolumnfalse}
]
\section{Introduction}

\begin{figure*}[t]
  \centering
  \includegraphics[width=\textwidth]{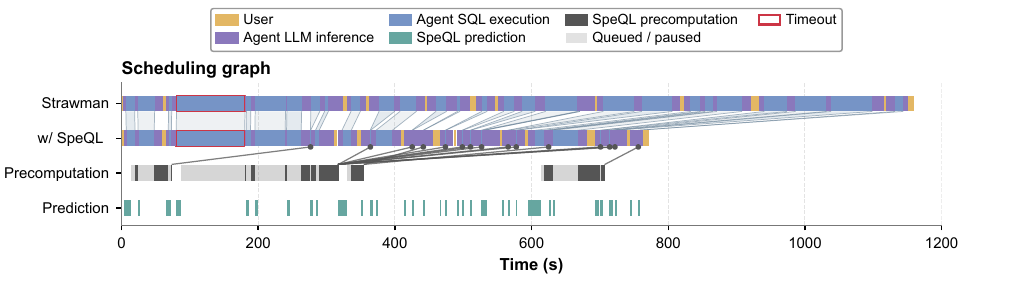}
  \caption{An exploration trace in which an agent investigates publication growth across computer science subfields. Ribbons connect corresponding SQL calls across systems; dark-gray links indicate partial result reuse.}
  \label{fig:task1-scheduling-resources}
\end{figure*}

{\em ``Wisdom begins in wonder.'' --- Socrates.}

Substantial effort and expense go into accelerating LLM inference, yet faster models still spend time waiting for tools. When an agent queries a database or invokes an external service, its next reasoning step often depends on the result. Tool execution can therefore consume much of the time saved by faster inference. Cortex reports that external retrieval accounts for approximately 40--50\% of execution time in its evaluated Search-R1 deployment~\cite{ruan2026cortex}. Tool calls also complicate inference memory management. During a tool invocation, retaining the agent's KV cache occupies GPU memory, while reclaiming that memory can introduce reloading, recomputation, and queueing delays when the agent resumes~\cite{li2025continuum}. Accelerating tool execution is therefore an important complement to accelerating model inference.

We focus on database analytics, where SQL queries form the interface between an LLM and the data it analyzes. Commercial systems such as Snowflake Cortex Analyst and Databricks Genie already support natural-language exploration of enterprise data~\cite{snowflakeCortexAnalyst,databricksGenie}. Consider an assistant investigating which computer science subfields are growing fastest using a publication dataset (Figure~\ref{fig:task1-scheduling-resources} Strawman). It compares publication trends, checks unexpected growth, and examines citations. Each query informs subsequent reasoning, placing database execution repeatedly on the critical path. In our agent workload, SQL execution accounts for approximately 71\% of the baseline's combined reasoning and active SQL time on the full dataset. Reducing these waits directly addresses a major component of the time users spend waiting for analysis.

The same latency problem arises when humans write SQL directly. Analysts alternate between constructing queries, inspecting results, and refining their hypotheses~\cite{chatzopoulou2009query}. Slow queries interrupt this process; a controlled study of interactive visual analysis found that adding 500\,ms per operation reduced user activity, dataset coverage, and the rate of hypothesis generation~\cite{liu2014latency}. Both agent-driven and human-driven analytics therefore present an opportunity: while the agent reasons or the analyst types, the database can prepare for queries that have not yet been submitted~\cite{polyzotis2003speculative,xin2021enhancing}. The challenge is that these queries are ad hoc. Their structure and parameters emerge during exploration, and later queries depend on information that may not yet be available.

We present \sys{}, a system that uses LLM-guided speculation to move database work ahead of query submission. In agent workflows, a small auxiliary LLM runs alongside the primary model and predicts its likely data needs. In a SQL editor, the speculator instead uses the analyst's partially written query. The central insight is that useful preparation does not require predicting an exact query: several plausible future queries may share a query shape, access the same data, or be answered from a common intermediate result. \sys{} exploits this overlap to make speculation useful even when the eventual query differs from the prediction.

\sys{} obtains benefits at three levels. First, executing speculative queries populates the database's existing planning and compilation caches, preparing query shapes that can be reused even when constants change~\cite{armenatzoglou2022amazon}. Second, speculative execution brings relevant data into memory. Third, \sys{} materializes temporary tables containing supersets of the data needed by anticipated queries. For example, if a future query is likely to filter sales by quantity but the threshold is uncertain, a temporary table can retain the quantity column and defer that filter. Multiple subsequent thresholds can then be evaluated over the same, smaller intermediate table. This approach targets reusable work across possible queries rather than requiring an exact result-cache hit~\cite{scully2017program}.

To realize this approach, \sys{} combines an LLM-guided speculator with a rule-based rewriter. The speculator predicts and prepares useful partial results, shown in the Prediction and Precomputation rows of Figure~\ref{fig:task1-scheduling-resources}. When SQL arrives, the rewriter identifies parts that can reuse these results and rewrites them using view matching~\cite{halevy2001answering}. A prediction can therefore help without matching the complete query. Background preparation runs while the agent reasons and yields whenever the agent submits SQL. Together, these techniques shorten the illustrated exploration session from 19.3 to 12.9 minutes.

The human-facing interface uses the same preparation mechanisms while analysts compose SQL. It additionally repairs incomplete query fragments and previews intermediate results, showing speculative edits explicitly so that users can interpret the displayed data. These previews can help analysts refine a query or discover an answer before completing the SQL they originally intended to write. Thus, the same speculative engine can both shorten database waits in agent workflows and support interactive exploration by human analysts.

\textbf{Limitations.}
\sys{} helps most when SQL execution takes up much of the user's wait. Its benefits are smaller when queries are already fast or most time is spent on reasoning or other tools. It also needs enough context to infer intent, time to prepare partial results, and a focus that stays consistent long enough to reuse them.

\textbf{This paper makes the following contributions:}
\begin{itemize}[leftmargin=1.1em]
    \item We identify an opportunity to accelerate SQL tool calling by predicting shared partial results that are easier to anticipate than complete queries yet account for much of their execution cost.
    \item We design and implement SpeQL, an LLM-guided system that predicts, precomputes, and reuses partial results through DAG-based query decomposition.
    \item We evaluate \sys{} in agent workflows and a 24-participant user study. Across 10 agent tasks with 384 SQL calls on a 771\,GB OpenAlex~\cite{priem2022openalex} dataset, \sys{} reduces mean per-task SQL execution time by 31.8\% and user-perceived latency by 22.6\%, with 6\% higher API cost and 10\% more physical read I/O.
    In a separate study, 24 participants completed two exploratory analysis tasks on 100\,GB of TPC-DS data hosted on Amazon Redshift. At a system cost of \$4 per hour, \sys{} reduces query latency on both tasks and significantly shortens time-to-answer on the second. Crucially, \sys{}'s previewed data helps user finds the results more quickly, and 87.5\% of participants reported improved productivity.
\end{itemize}

\section{\sys{}}
\label{sec:system-design}

\sys{} supports both agents issuing SQL and humans incrementally writing SQL in an editor. Both settings offer time to prepare useful work before a query is submitted. To keep the section focused, we first describe the agent workflow, then expand the design to human-written SQL and live result previews.

\subsection{Preparing reusable work}

Consider a user asking an AI agent: ``We are planning a back-to-school campaign. Which products should we promote?'' The agent has checked overall monthly revenue and is deciding what to examine next. Analyzing a large sales history takes time. Faster model reasoning still leaves the agent waiting for SQL. Can we prepare useful database work while the agent reasons?

\para{Speculation needs an advantage.}
Effective speculation exploits some structure that lets a predictor get ahead. Hardware predictors make quick guesses while the processor waits for conditions or data~\cite{smith1981branch,jouppi1990cache}. Speculative decoding uses a fast draft model to advance generation~\cite{leviathan2023fast}. For SQL tool calls, however, asking an LLM to predict the agent's next query provides no automatic advantage. It must resolve much of the same uncertainty, then leave enough time to execute its prediction. We need a prediction task that is easier than deciding the complete analysis.

\para{Predicting exact queries is difficult.}
A natural approach is to predict and execute multiple possible next queries~\cite{ye2026speculativeactions}. This increases the chance of preparing a useful result, but also increases contention for database resources. Even then, exact predictions remain brittle. For example, one might predict ``Find the ten bestselling backpacks during last year's back-to-school season,'' while the agent actually compares sales across school-related product categories. The backpack ranking cannot answer the category question. In our evaluation, exact-query speculation successfully precomputes 74 query results, yet none is reused across 384 agent SQL calls.

\begin{figure}[htpb]
  \centering
  \includegraphics[width=\linewidth]{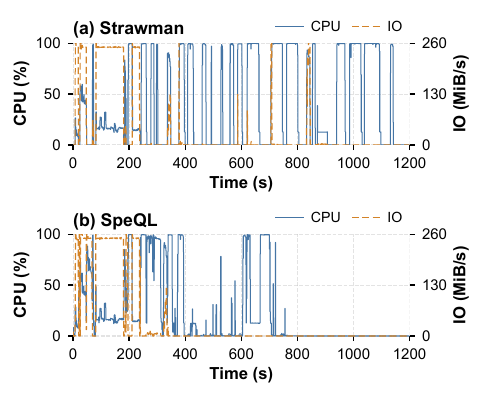}
  \caption{CPU utilization and physical read throughput during an exploration session under (a) Strawman, the baseline without \sys{} and (b) \sys{}.}
  \label{fig:task1-cpu-io-stacked}
\end{figure}

\para{Preparing everything is expensive.}
An alternative is to prepare data that cover all possible future queries. We could compute sales for every grouping of products, categories, customers, and stores over every date range. This covers both the product ranking and category comparison, but the prepared results can exceed the original sales history in size and make subsequent queries slower. Simply fetching raw data into memory avoids this explosion, but leaves the work of processing those data to future queries. Figure~\ref{fig:task1-cpu-io-stacked}(a) shows resource usage for the Strawman execution in Figure~\ref{fig:task1-scheduling-resources}. In the later stages, CPU utilization remains high despite little physical I/O.  Bringing data into memory alone does not eliminate these processing costs.

\para{Opportunity: different questions share preparation.}
In the back-to-school example, both queries draw on sales of school-related products during the previous back-to-school season. Although the user has not named specific products, the campaign context allows the LLM to infer likely relevant categories, such as backpacks and stationery. We can therefore use the conversation to prepare relevant sales records in advance and let the agent decide how to analyze them. This preparation can dominate query time because it processes a large sales history.

\para{Insight: preparing shared work.}
The overlap between the predicted and actual queries reflects a common pattern in exploration. Users and agents ask different questions while remaining focused on a bounded data scope. Their questions evolve, but some expensive preparation is repeatedly needed. Preparing these results can be both easier to anticipate than the complete query and responsible for much of its execution cost.

\begin{figure}[htpb]
  \centering
  \includegraphics[width=\linewidth]{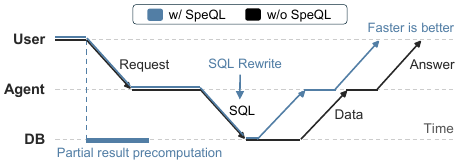}
  \caption{\sys{} uses the conversation to predict and precompute reusable partial results. When the agent issues SQL, \sys{} rewrites it to reuse the prepared data. Completing shared work in advance allows the query to return results more quickly. }
  \label{fig:wonder-workflow}
\end{figure}

\sys{} exploits this structure by predicting shared \emph{partial results}. \sys{} prepares these results while the agent reasons, then rewrites incoming SQL to reuse them and run faster (Figure~\ref{fig:wonder-workflow}). Unlike exact-query speculation, a partial result can remain useful when the agent chooses a different analysis. Unlike data prefetching, it saves repeated processing rather than merely bringing raw data into memory. For this approach to work, partial results must be easier to predict than complete queries, while their preparation accounts for much of the queries' execution cost. We find that these two properties can coexist in interactive data exploration.

We find this simple insight effective. In our evaluation, prompting the LLM to predict shared partial results instead of complete future queries raises the SQL match rate from zero to about 30\% across 384 calls. For matched calls, median execution time falls from 31.87 seconds to 192 milliseconds. Figure~\ref{fig:task1-cpu-io-stacked}(b) also shows that \sys{} saves CPU work rather than merely preloading data. Together, these results show that shared work is easier to predict, and preparing it in advance can save much of a query's execution time.

\subsection{System Design}

\sys{} has two components: an LLM-guided \emph{speculator} and a rule-based \emph{rewriter}. The speculator uses the conversation to predict and prepare useful partial results. When SQL arrives, the rewriter modifies it to reuse these results.

\begin{figure*}[htbp]
  \centering
  \includegraphics[width=\linewidth]{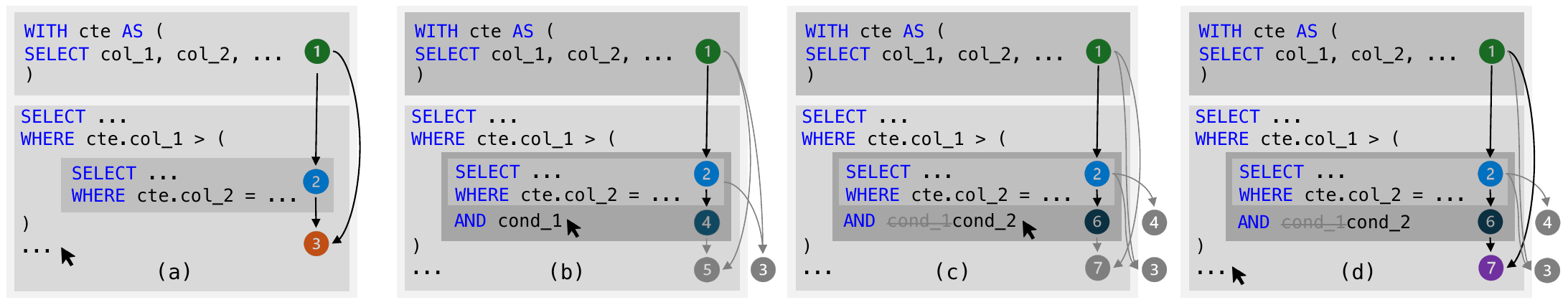}
  \caption{\sys{} supports interactive query editing with reusable partial results and live previews. Each node represents a \texttt{SELECT} statement; colored nodes have precomputed results.}
  \label{fig:query-editing-workflow}
\end{figure*}

\subsubsection{Speculator: preparing partial results}
\label{sec:speculator}

\para{Predict from intent.}
The speculator uses the user-agent interaction history to infer what data may be needed next. This history includes user requests, agent outputs, SQL queries and their results, all with timestamps. The speculator runs asynchronously alongside the agent.

The speculator runs whenever a user submits a request or the agent issues SQL. These events provide complementary signals: the request reveals what the user wants to investigate, while the SQL reveals how the agent is approaching it. Each event prompts a small LLM to choose useful preparation for subsequent queries rather than predict a complete future answer.

Alongside the task instruction, the prompt includes the database schema, the available interaction history, completed partial results, and the query currently being materialized, if any. Completed results help the model avoid preparing the same data again, while the ongoing query lets it decide whether to preserve earlier work or redirect resources.

The model returns one of three decisions:
\begin{itemize}
    \item \texttt{KEEP}: continue the current preparation and start no new query.
    \item \texttt{CANCEL}: stop the current preparation without replacing it.
    \item \texttt{EXECUTE}, accompanied by executable SQL: stop the current preparation, if any, and run a new query to cache a partial result (using \sql{CREATE} \texttt{TEMPORARY TABLE}~\cite{ben2000temporary}).
\end{itemize}
These decisions let \sys{} keep useful work and redirect preparation as the conversation evolves.

\para{Prioritize requested SQL.}
Background preparation yields to agent SQL and uses the time between calls. Prioritizing preparation may help future queries, but can delay the agent's current answer. The agent therefore reuses completed results without waiting for unfinished preparation.

\subsubsection{Rewriter: reusing partial results}
\label{sec:rewriter}

The rewriter finds parts of an incoming query that can reuse completed partial results. In the back-to-school example, the agent can reuse the prepared sales data to compare categories or find popular products, without repeating the shared work.

\para{Find reusable parts.}\label{sec:dag-construction}
To find reusable work within a query, the rewriter breaks it into smaller parts. Many SQL queries naturally consist of several \texttt{SELECT} statements; for example, queries in our workload contain up to 17. The rewriter organizes these statements into a directed acyclic graph (DAG). Each node represents a \texttt{SELECT} statement, and an edge shows that one statement uses another's result. This lets the rewriter match partial results to individual parts without predicting the complete query.

\para{Match and rewrite.}
For each node, the rewriter checks completed partial results from newest to oldest. It compares their parsed structure~\cite{mao2024sqlglot} to see whether a prepared result contains the information needed for that part of the query. When a match is found, the rewriter replaces the shared work with the prepared result and keeps any remaining operations~\cite{halevy2001answering,zaharioudakis2000answering}. If no equivalent rewrite is found, it falls back to the original SQL. These rule-based rewrites preserve result equivalence, require no additional LLM calls, and take only a few milliseconds. Appendix~\ref{app:rewriter-rules} details the rewrite rules.

\subsubsection{Extension to human-written SQL}
\label{sec:human-written-sql}
\label{sec:workflow-example}

\sys{} also supports users writing SQL in an editor, inspecting results and refining the query along the way.

\para{Example: edit, preview, and refine.}
Figure~\ref{fig:query-editing-workflow}(a) shows a query with three parts: an initial calculation~\ding{172}, a subquery~\ding{173} that uses its result, and a main query~\ding{174}. The user is editing the main query, which is still incomplete and cannot execute. \sys{} repairs the input, prepares reusable partial results, and previews the main query's result. The user can inspect the data before finishing the SQL.

The preview helps the user notice that the subquery needs another condition. The user moves the cursor to the subquery and adds \texttt{cond\_1}, producing query~\ding{175} in Figure~\ref{fig:query-editing-workflow}(b). \sys{} applies this condition to the prepared result of~\ding{173} and quickly updates the preview.

The user then changes \texttt{cond\_1} to \texttt{cond\_2}, producing query~\ding{177} in Figure~\ref{fig:query-editing-workflow}(c). The result of~\ding{175} may no longer contain the needed data, but the earlier result of~\ding{173} still does. \sys{} reuses it to compute~\ding{177} and update the preview.

Finally, the user returns to the main query in Figure~\ref{fig:query-editing-workflow}(d). The subquery edits have changed it from~\ding{174} to~\ding{178}. \sys{} computes the updated query using the prepared results of~\ding{172} and~\ding{177}, avoiding repeated work over the original data.

\para{Design goal: quick feedback during editing.}
As the example illustrates, beyond accelerating submitted queries, we want users to see results as they change numbers, columns, and conditions, even when the SQL is incomplete. Ideally, exploration should feel like editing a spreadsheet: a small change quickly reveals its effect. This requires shortening both LLM inference for repair and speculation and SQL execution for previews.

\para{Reuse repair patches across edits.}
Users edit SQL much more frequently than agents submit queries. Generating a complete query after every small edit would be slow and costly. Yet users usually change one part at a time, so the same repair can remain useful across several edits.

\sys{} therefore caches LLM-generated repairs~\cite{chen2023teaching} as patches~\cite{Diff-match-patch}. When the input changes, it first applies the cached patch. If the repaired query is valid, it skips another LLM call. Otherwise, the LLM generates a new patch rather than the entire query. This saves both LLM calls and generation time.

\begin{figure}[htpb]
  \centering
  \includegraphics[width=\linewidth,trim=0 12 0 60,clip]{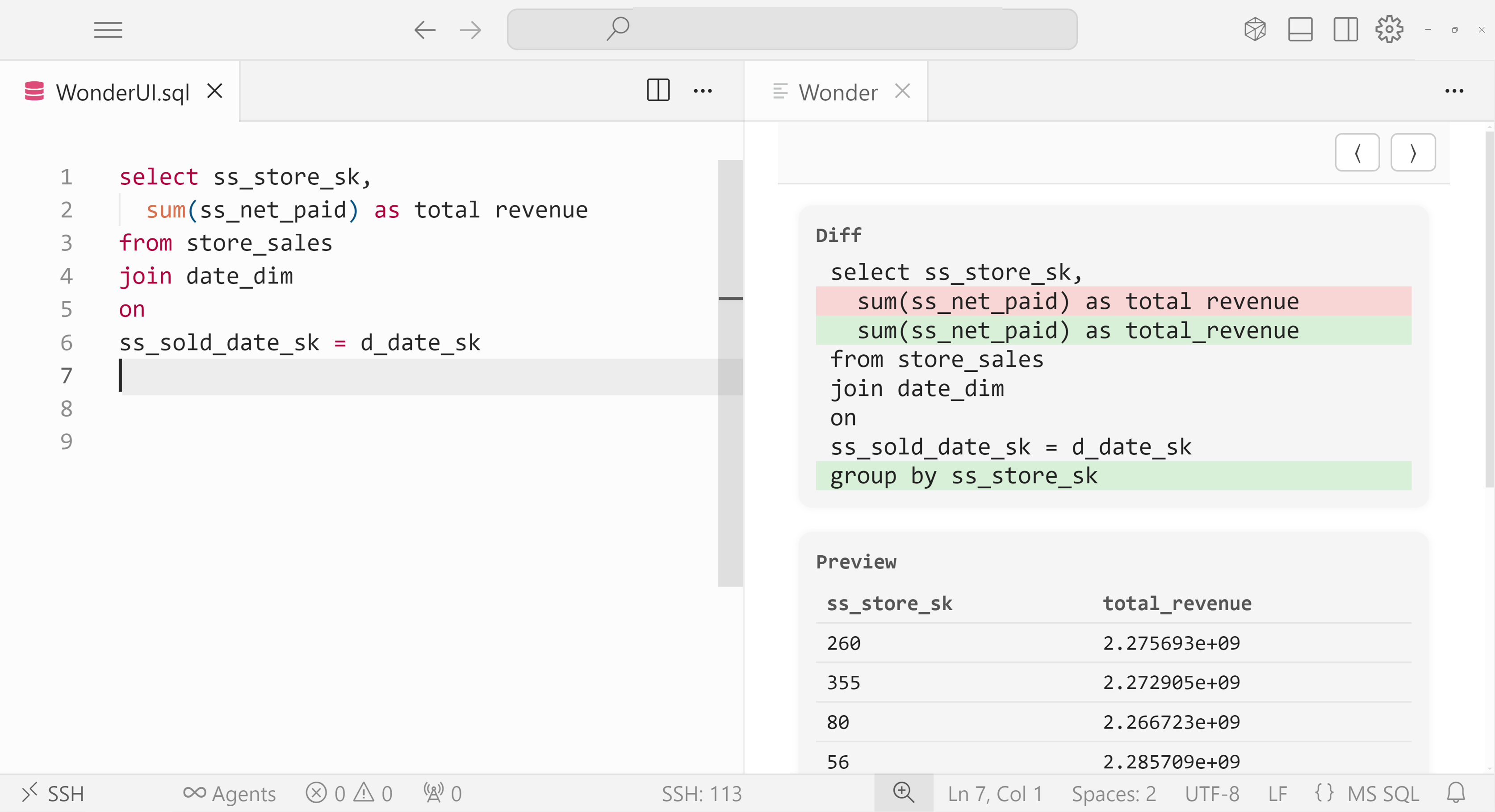}
  \caption{\sys{} in a SQL editor. The left panel shows the user's incomplete query; the right panel shows the inferred changes and result preview.}
  \label{fig:wonder-interface}
\end{figure}

\para{Bound speculation using over-projection.}
Unlike an agent conversation, an editor may provide little explanation of the user's intent beyond the SQL itself. To focus LLM speculation on what the user is working on, \sys{} uses the cursor position to identify the statement being edited. Since users often continue editing the same statement, \sys{} predicts likely additions to that statement.

To keep speculation focused and inference fast, \sys{} proposes \emph{over-projection}. It asks the LLM to predict a few additional columns the user may need, then adds them to the statement's \texttt{SELECT} list, adjusting \texttt{GROUP BY} when needed. This prepares data for later edits while requiring the LLM to generate only column names.

\para{Reuse partial results for live previews.}
The rewriter reuses completed partial results to accelerate the repaired query and update its preview. The same prepared data can support different conditions, such as \texttt{cond\_1} and \texttt{cond\_2} in the example. Users can therefore inspect results and refine their queries with less waiting. The editor also shows the inferred changes as a diff, so users can see what was executed (Figure~\ref{fig:wonder-interface} shows \sys{}-supported SQL editor).

\section{Evaluation}\label{sec:evaluation}

\para{Goal}
We evaluate whether \sys{} accelerates data exploration by preparing reusable partial results before queries arrive. We study both agents issuing SQL and users writing SQL interactively. Our evaluation answers three questions:

(1) How much does reusing partial results reduce latency throughout data exploration?

(2) What resources and monetary costs does this acceleration require?

(3) How does \sys{} improve users' exploration efficiency and experience?

We also use TPC-DS benchmarks~\cite{tpcds} to test \sys{}'s DAG-based rewriting on complex queries with multiple joins, filters, and nested expressions. We report these experiments in Appendix~\ref{app:microbenchmarks}.

\para{Datasets.}
Our evaluation uses two workloads.

\textbf{(1) Agentic exploration traces.}
We generate ten multi-turn exploration traces over OpenAlex~\cite{priem2022openalex}, a 771\,GB database of scholarly publications. Each trace involves a simulated user and a database-connected agent. The user LLM receives a private analysis goal, generated by Sonnet 5.5 and hidden from the agent. It pursues this goal through approximately ten rounds of requests, adapting to the agent's previous answers. The agent investigates each request through SQL tool calls, with a timeout of up to 100 seconds per call.

We use the user LLM's inference time to represent human thinking time and add no idle time. Longer pauses could provide more time for preparation. The private goals and LLM prompts are available in Appendix~\ref{app:agentic-prompts}. Neither the user nor the agent is prompted to favor queries that enable partial-result reuse.

\textbf{(2) User study.}
We collect real typing data through a utility/user study on the 100\,GB TPC-DS dataset~\cite{tpcds}. Participants answer two newly designed analysis questions that require them to iteratively investigate metrics and trends in store revenue. We record their input behavior, query latency, and task completion time. The study captures how users construct, debug, and refine SQL during interactive exploration.

\para{Metrics.}
For agentic workflows, we measure user-perceived latency per round, SQL tool-call time, and CPU, I/O, and API usage. For the user study, we measure user-perceived query latency, time-to-answer, database and API cost, and subjective ratings.

\para{Database.}
For agentic workflows, we run DuckDB~\cite{raasveldt2019duckdb} on an AWS EC2 m8i.24xlarge instance with 48 CPUs and 372\,GiB of RAM. We limit the workflow to eight CPUs without imposing a memory limit. This represents a cloud deployment with a fixed compute quota and a shared memory pool.

For the user study, we use Amazon Redshift Serverless~\cite{armenatzoglou2022amazon} with a maximum of eight RPUs. Redshift Spectrum~\cite{awsRedshiftSpectrum} queries TPC-DS data stored in Amazon S3~\cite{TPCDSData}, allowing us to evaluate \sys{} on a commercial database without modifying its execution engine.

\para{LLMs.}
For agentic workflows, \sys{} uses Haiku 5.5 to predict partial results, while both the simulated user and the agent use Sonnet 5.5. For the user study, we use GPT-4o-2024-08-06 and GPT-4o-mini.

\para{Baselines.}
For agentic workflows, we compare \sys{} with three baselines on the same recorded traces: (1)~\textit{Strawman}, which executes the agent's SQL without speculation;
(2)~\textit{Speculative Actions}~\cite{ye2026speculativeactions}, which predicts and executes complete future SQL tool calls, reusing their results only when the predicted and subsequent queries are equivalent. It uses the same scheduler as \sys{};
(3)~\textit{Rule-based MV}, our implementation based on AutoAdmin's view merging~\cite{agrawal2000automated}, which precomputes shared results from past queries with a 100-second active-time budget per build. It uses the same rewriter and scheduler as \sys{}, but selects precomputation candidates using fixed rules over historical SQL rather than LLM predictions informed by conversational context.
For the user study, the baseline uses the same Redshift endpoint without \sys{}'s speculative execution or interactive previews.

\begin{table*}[t]
\centering
\caption{Agentic workflow measurements. All non-count metrics are means over 10 tasks. Percentage changes are relative to Strawman.}\label{tab:agentic-workflow-metrics}
\small
\setlength{\tabcolsep}{3pt}
\begin{tabularx}{\textwidth}{@{}Xlrrrr@{}}
\toprule
\textbf{Per-task Metric} & \textbf{Scale} & \textbf{Strawman} & \textbf{Speculative Actions} & \textbf{Rule-based MV} & \textbf{SpeQL} \\
\midrule
User-perceived time (s) $\downarrow$ & Full & 1842.78 & 1839.44 (-0.18\%) & 1733.02 (-5.96\%) & \textbf{1426.85 (-22.57\%)} \\
 & 1/4 & 829.42 & 826.55 (-0.35\%) & 773.83 (-6.70\%) & \textbf{737.15 (-11.12\%)} \\
 & 1/16 & 617.37 & 615.28 (-0.34\%) & 604.48 (-2.09\%) & \textbf{594.17 (-3.76\%)} \\
\midrule
SQL execution time (s) $\downarrow$ & Full & 1307.11 & 1303.77 (-0.26\%) & 1197.35 (-8.40\%) & \textbf{891.18 (-31.82\%)} \\
 & 1/4 & 293.75 & 290.88 (-0.98\%) & 238.15 (-18.93\%) & \textbf{201.48 (-31.41\%)} \\
 & 1/16 & 81.70 & 79.60 (-2.56\%) & 68.81 (-15.78\%) & \textbf{58.50 (-28.40\%)} \\
\midrule
Matched SQL / All SQL $\uparrow$ & Full & 0 / 384 & 0 / 384 & 51 / 384 & \textbf{115 / 384 (86 are DAG nodes)} \\
 & 1/4 & 0 / 384 & 0 / 384 & 93 / 384 & \textbf{131 / 384 (104 are DAG nodes)} \\
 & 1/16 & 0 / 384 & 0 / 384 & 77 / 384 & \textbf{109 / 384 (83 are DAG nodes)} \\
\midrule
Timed-out SQL / All SQL $\downarrow$ & Full & 67 / 384 & 67 / 384 & 67 / 384 & \textbf{50 / 384} \\
 & 1/4 & 11 / 384 & 10 / 384 & 10 / 384 & \textbf{9 / 384} \\
 & 1/16 & \textbf{0 / 384} & \textbf{0 / 384} & \textbf{0 / 384} & \textbf{0 / 384} \\
\midrule
CPU time (core$\cdot$s) $\downarrow$ & Full & 6762.86 & 8485.85 (+25.48\%) & 9391.47 (+38.87\%) & \textbf{5247.36 (-22.41\%)} \\
 & 1/4 & 1162.79 & 1636.47 (+40.74\%) & 2758.54 (+137.23\%) & \textbf{827.29 (-28.85\%)} \\
 & 1/16 & 342.38 & 474.72 (+38.65\%) & 843.11 (+146.25\%) & \textbf{235.49 (-31.22\%)} \\
\midrule
I/O read (GiB) $\downarrow$ & Full & \textbf{74.85} & 78.10 (+4.35\%) & 95.43 (+27.50\%) & 82.06 (+9.65\%) \\
 & 1/4 & \textbf{19.08} & 21.72 (+13.83\%) & 34.79 (+82.32\%) & 24.08 (+26.18\%) \\
 & 1/16 & \textbf{4.58} & 5.00 (+8.96\%) & 9.18 (+100.15\%) & 5.68 (+23.96\%) \\
\midrule
API cost (USD) $\downarrow$ & All & \textbf{1.3150} & 1.4414 (+9.62\%) & \textbf{1.3150 (0.00\%)} & 1.3919 (+5.85\%) \\
\midrule
Peak DRAM (GiB) $\downarrow$ & Full & \textbf{19.29 / 86.94} & 28.76 / 98.64 & 60.06 / 146.34 & 22.57 / 102.56 \\
anonymous / total & 1/4 & \textbf{26.64 / 42.50} & 37.80 / 57.33 & 53.88 / 100.09 & 31.73 / 54.53 \\
 & 1/16 & \textbf{7.72 / 11.83} & 13.93 / 18.21 & 19.27 / 32.81 & 8.37 / 13.48 \\
\bottomrule
\end{tabularx}
\end{table*}

\subsection{Agentic Workflow}
\label{sec}

\subsubsection{Setup}
We evaluate how \sys{} accelerates an agent's analytical SQL tool calls and reduces user-perceived latency. We replay the recorded agentic traces, holding agent reasoning durations fixed while allowing SQL execution times to vary across \sys{} and the baselines.  This controlled comparison isolates the latency effects of speculative execution. Our later user study examines how \sys{} affects query behavior during live interaction. For every system that rewrites queries, we verify that each rewritten query produces the same result as the original.

\begin{figure}[t]
  \centering
  \includegraphics[width=\linewidth]{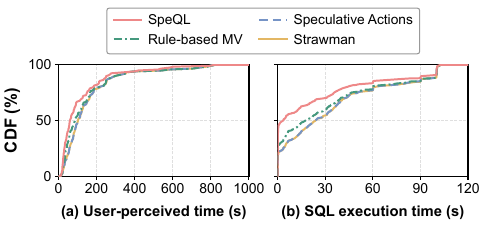}
  \caption{Latency CDF for \sys{} and baselines.}
  \label{fig:agentic-workflow-latency-cdfs}
\end{figure}

\subsubsection{Results}

\para{Latency reduction.}
\sys{} reduces average SQL execution time by 31.82\%, 31.41\%, and 28.40\% on the Full, 1/4, and 1/16 datasets, respectively (Table~\ref{tab:agentic-workflow-metrics}), relative to Strawman. This reduces average user-perceived latency by 22.57\%, 11.12\%, and 3.76\%. The larger datasets see greater user-perceived gains because SQL execution accounts for more of the response time. On the Full dataset, 44.01\% of SQL calls complete within one second with \sys{}, compared with 18.49\% with Strawman, highlighting the effectiveness of partial-result reuse (Figure~\ref{fig:agentic-workflow-latency-cdfs}). \sys{} also outperforms Rule-based MV, which identifies shared computation from past SQL using fixed rules but cannot use user or agent intent to anticipate future queries.

Across the three scales, 29.95\%, 34.11\%, and 28.39\% of SQL calls reuse a precomputed partial result. Most matched queries incorporate the result through a CTE, subquery, or join, showing how DAG-based rewriting supports reuse within complex queries. In contrast, Speculative Actions obtains no exact speculation hits and reduces user-perceived latency by less than 0.4\%. Partial results therefore support queries whose complete SQL is difficult to predict. On the Full dataset, \sys{} also reduces timed-out calls from 67 to 50, allowing more queries to finish within the timeout budget.

\para{Resource and API cost.}
\sys{} saves CPU time while adding moderate I/O, memory, and API costs. Including partial-result preparation, total CPU time decreases by 22.41--31.22\% relative to Strawman, as reuse avoids repeated computation across queries. Rule-based MV also reuses partial results, but its preparation costs outweigh the CPU savings.  For \sys{}, physical reads increase by 9.65--26.18\%, and peak total DRAM occupancy increases by 13.91--28.30\%. Partial-result prediction uses the inexpensive Haiku 5.5 model, costing an average of $\$0.077$ per task.

\begin{figure}[t]
  \centering
  \includegraphics[width=\linewidth]{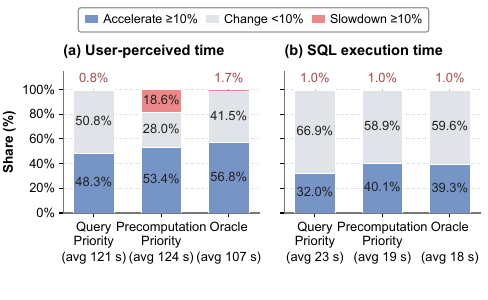}
  \caption{Scheduling policy comparison. Bars show the share of (a) rounds and (b) SQL calls with lower, similar, or higher latency relative to Strawman.}
  \label{fig:agentic-scheduling-policy-comparison}
\end{figure}

\para{Tradeoff between precomputation and responsiveness.}
\sys{} yields precomputation to agent SQL to avoid delaying tool calls. However, yielding can leave useful precomputation unfinished when queries arrive. Figure~\ref{fig:agentic-scheduling-policy-comparison} examines this tradeoff on the Full dataset. It compares \textit{Query Priority}, used by \sys{}, with \textit{Precomputation Priority}, which completes ongoing generation before running pending agent SQL, and Oracle, which uses future trace information to choose whether generation should block a query.

Precomputation Priority reduces mean SQL execution time by 19.90\% relative to Query Priority, showing the benefit of completing precomputation earlier. However, 18.6\% of rounds experience a user-perceived slowdown of at least 10\% relative to Strawman, compared with only 0.8\% under \sys{}. These delays offset the SQL savings, leaving Precomputation Priority with 2.37\% higher mean user-perceived latency than Query Priority. Both policies remain slower than Oracle. \sys{} adopts Query Priority because it delivers more consistent gains across rounds while keeping slowdowns uncommon.

\subsection{Utility/user study}\label{sec:utility-and-user-study}
\subsubsection{Methodology} To evaluate \sys{}'s improvements for human-written SQL, we implemented \sys{} as a VS Code SQL Editor plugin and recruited 24 participants to complete a 60-minute questionnaire\footnote{Participants and data preprocessing details are shown in Appendix~\ref{app:user-study-details}.}. They were instructed to connect to our server located in Utah, which further communicates with Azure OpenAI API services and a Redshift endpoint (with TPC-DS \cite{tpcds} 100GB data). 

The questionnaire includes two parts. \textbf{The first part measures the utility}, where we create two data analysis tasks (Q1 and Q2, as described in \S \ref{sec:utility-study}), based on the 100 GB TPC-DS data. We asked participants to write SQL and finish the two tasks as quickly as possible, during which we recorded their input behavior, query latency, and time-to-answer. To compare the utility improvement, participants were randomly assigned to two groups, A and B, without their knowledge: For group A, we instructed them to learn \sys{} before showing the tasks; For group B, we only informed them that we want to record their input behavior but did not introduce \sys{}. Both groups were given five minutes to run example SQL queries and recall SQL rules to familiarize themselves with the testbed. We applied the Mann-Whitney U test~\cite{mann1947test} to determine whether \sys{} significantly reduced task completion time, using a predefined significance level of $\alpha=0.05$, and used Cliff's Delta~\cite{cliff1993dominance} to quantify the effect size.\footnote{$p<0.05$: significant; $p\ge0.05$: insignificant; $|\delta|<0.147$: almost no difference; $0.147<|\delta|<0.330$: small difference; $0.330<|\delta|<0.474$: moderate difference; $|\delta|>0.474$: strong difference.}

\textbf{The second part measures the usability}, where participants fill in a standard system usability scale (SUS \cite{brooke1996sus}) questionnaire and leave open-ended comments on \sys{} according to their subjective experiences. Before rating the scores, group B participants undergo a debriefing session and are instructed to learn \sys{} and use it to finish the two utility tasks again.

\subsubsection{Utility study}\label{sec:utility-study}

We design two questions for participants that necessitate interacting with the data to derive the correct answers. Attempting to write a single SQL query without exploring the data will result in errors due to the presence of missing (\sql{NULL}) values or incomplete data. Participants were told to only consider eight columns.

\noindent\textbf{Q1.} "For different physical stores recorded in 2001, what is the highest annual revenue? Note, if you find the answer is close to $1.11 \times 10^9$ USD, you may probably make a common mistake."

\textbf{General querying pattern for Q1.} Nearly all participants initially listed the revenue for each store key without verifying whether the keys corresponded to valid stores, as anticipated. This led them to produce incorrect answers. Participants had to independently debug their queries to filter out invalid store keys (by checking an additional \texttt{ss\_store\_sk} column, or including an \texttt{ss\_store\_sk \sql{IS NOT NULL}} condition) before arriving at the correct result.

\begin{figure}[htbp]
  \centering
  \includegraphics[width=\linewidth]{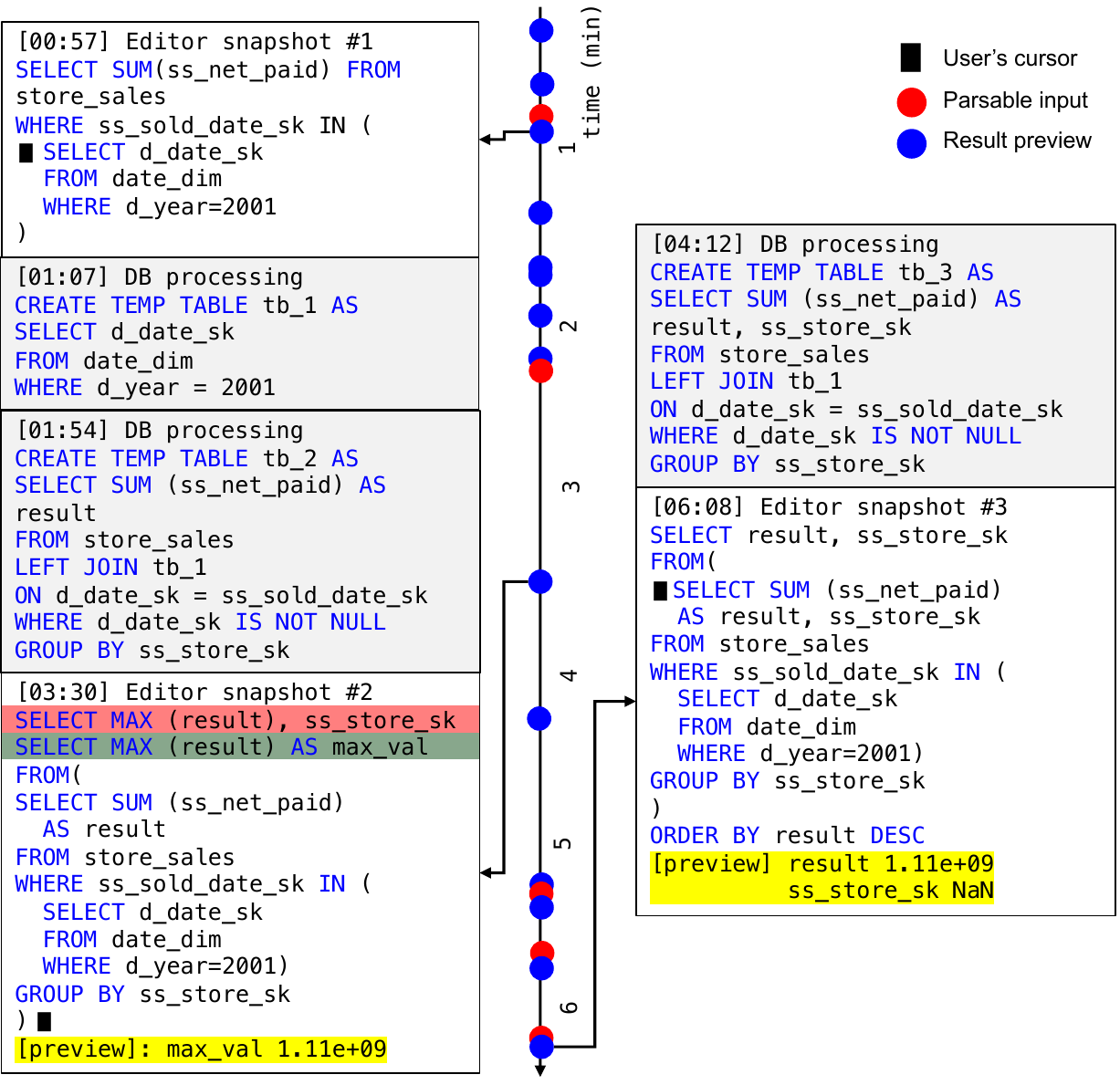}
  \caption{Case study for Q1.}
  \label{fig:user-study-query-timeline}
\end{figure}

\textbf{Case study for Q1.} We showcase a data point from Q1 to illustrate \sys{}'s workflow and how it helps reduce latency, as shown in Fig. \ref{fig:user-study-query-timeline}. The participant completed the task in approximately six minutes, which includes time spent thinking, typing, and debugging. From their input behavior, we derive two key observations, which are also observed in other participants: \textbf{(1) The query processing time often largely overlaps with the thinking time.} While typing, \sys{} created temporary tables quite early, specifically, \texttt{tb\_1}, \texttt{tb\_2} and \texttt{tb\_3} were created at 01:07, 01:54 and 04:12, respectively. This indicates that there would be sufficient time to generate temporary tables in advance. 
\textbf{(2) \sys{} reduces query latency by up to 289 times.} As we observed, the participant had a tendency to first execute simple SQL queries, inspect their results, and then integrate these into a larger, more complex logic. This behavior reinforces the feasibility of \sys{} materializing the results of subqueries, such as \texttt{tb\_1} in snapshot \#1, \texttt{tb\_2} in snapshot \#2, and \texttt{tb\_3} in snapshot \#3. With the help of the precomputed temporary tables, \sys{} completed the final query within 0.08 seconds, which is $289\times$ faster than the baseline (simply pasting and running the query in Redshift query editor takes 23.1 seconds). Furthermore, during the editing process, \sys{} provided nine previews (the blue circles), six of which occurred between the first and fifth minute when the SQL typically contained syntax errors. The participant later acknowledged that these previews were highly helpful for debugging and refining the SQL query.

\noindent\textbf{Q2.} "Calculate the total revenue for physical stores (sum of all \texttt{ss\_net\_paid} in the \texttt{store\_sales} table) for each year from 2000 to 2003. Analyze the reasons for the changes in total revenue in 2003."

\textbf{General querying pattern for Q2.} To address this question, participants first computed the total revenue across different years. They quickly observed that revenue in 2003 was significantly lower. By crafting more advanced queries, possibly after checking quantities and number of stores, they eventually discovered that the data for 2003 was truncated starting from January and the dataset is incomplete.

\textbf{Case study for Q2.} One particularly interesting finding in Q2 is \textbf{\sys{} helps users stop querying earlier once they find the desired information.} For instance, to calculate the total revenue, participants in group B typically wrote the following query (with result in the comment):

\noindent{\ttfamily \sql{SELECT} d\_year, \sql{SUM}  (ss\_net\_paid)} \noindent{\ttfamily \sql{FROM} store\_sales \sql{JOIN}}
\noindent{\ttfamily  date\_dim \\ \sql{ON} ss\_sold\_date\_sk = d\_date\_sk}
\noindent{\ttfamily \sql{WHERE} d\_year >= 2000 \sql{AND} d\_year <= 2003}\\
\noindent{\ttfamily \sql{GROUP BY} d\_year} \noindent{\ttfamily \sql{ORDER BY} d\_year;}\\
\noindent{\ttfamily {/* 2000: 9e10; 2001: 9e10; 2002: 9e10; 2003: 1e9 */}}
\\

\noindent This query is entirely correct, as it applies strict filtering conditions to eliminate  out-of-bounds data. However, participants often needed more time to write such a query, and most of them remained confused about the revenue drop for a while. In contrast, we observed that participants in group A derived the same insights much faster. They typed (with gray code generated by LLMs):

\noindent{\ttfamily \sql{SELECT} d\_year, \sql{SUM} (ss\_net\_paid)} \noindent{\ttfamily \sql{FROM} store\_sales \sql{JOIN} date\_dim}\\
\noindent{\ttfamily \rule[-0.3ex]{0.5pt}{2ex}\textcolor{grey}{ON ss\_sold\_date\_sk = d\_date\_sk GROUP BY d\_year}}\\
\noindent{\ttfamily {/* 2002: 9e10; 2003: 1e9; 1998: 9e10; 1999: 9e10; 2000: 9e10; 2001: 9e10 */}}
\\

\noindent The preview included additional years of revenue that users did not initially intend to request. However, it unexpectedly helped them identify the time boundary: 2003 was the last year in the dataset. As a result, we observed that participants in group A quickly realized the data was incomplete. We did not anticipate this benefit when designing the tasks.

\begin{figure}[t]
  \centering
  \includegraphics[width=\linewidth]{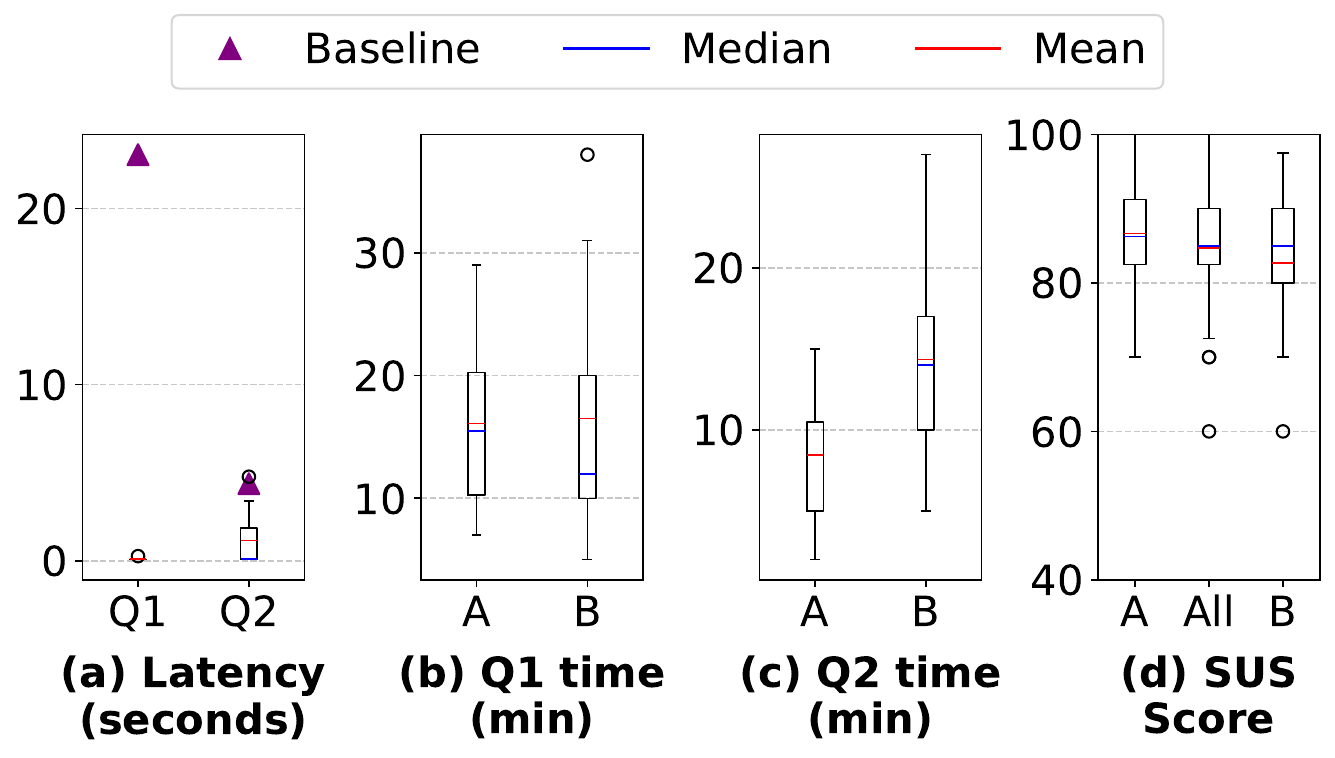}
  \caption{Utilily/user study. In (a), the boxes represent \sys{}'s latency while the purple triangles represent the baseline's. In (b), (c), (d), the boxes in group A are \sys{}'s result while that in group B are the baseline's.}
  \label{fig:user-study-results}
\end{figure}

\textbf{Statistics.}
(1) Fig. \ref{fig:user-study-results} (a) presents the query latency measured immediately after group A participants composed and finalized the runnable queries in their editor. In contrast, the baseline represents the query latency observed when the queries were directly issued without any preprocessing. We find that \textbf{\sys{} significantly and effectively reduces latency}, with a significance level of $p<0.001$, and Cliff's Delta $\delta=-1.0$ for Q1, and $p<0.001, \delta=-0.778$ for Q2. (2) Figure \ref{fig:user-study-results} (b) depicts the overall time-to-answer of Q1, where we observed neither significance ($p=0.775$) nor difference ($\delta=0.089$) between the two groups. We guess that this could be a limitation of our utility study setup---we should have created a warm-up task Q0 to help them get accustomed to \sys{} and the dataset. (3) Figure \ref{fig:user-study-results} (c) presents the overall time-to-answer of Q2, demonstrating that \textbf{\sys{} significantly ($p=0.025$) and effectively ($\delta=-0.570$) improves participants' time-to-answer}. We observe \sys{}'s previews provide additional information while they are typing, which helps them quickly eliminate unlikely exploration directions. (4) As a cost, each group A participant spent about \$0.5 on LLM API (not shown in the figure) and \$1.5 on database usage.

\subsubsection{User study}\label{sec:usability-study}
This section captures users' \textit{subjective} perceptions of \sys{}'s usability and utility. \textbf{The System Usability Scale (SUS \cite{brooke1996sus}) confirms the high usability of \sys{}.} As shown in Fig.~\ref{fig:user-study-results}, SUS scores between the two groups showed no significant difference ($p=0.771$, $\delta=-0.076$), indicating both found \sys{} valuable. \sys{} achieved an excellent SUS score of 85.3, including a learnability score of 85.8 and usability score of 83.3. This score outperforms approximately 95\% of systems, as illustrated in Fig. \ref{fig:user-study-results} (d), indicating that users are happy to use, and are willing to promote \sys{} in production. Notably, 58.3\% said they'd use \sys{} frequently for SQL, and 79.2\% strongly agreed it's easy to learn. \textbf{The open-ended survey reveals strong productivity gains and areas for improvement in \sys{}.} 87.5\% of participants reported increased productivity, noting that \sys{} made writing SQL and exploring data easier.

\section{Related work}

Speculative query processing \cite{polyzotis2003speculative} was first introduced in 2003, revealing the opportunity to overlap query execution with user input.  However, it relied on predefined query templates, where relations were represented in a tabular format. Users were required to construct queries by manually placing projection indicators and selection predicates on the corresponding fields. This assumption falls short in ad-hoc queries, where users often compose complex and structurally diverse queries \cite{hatami2020understanding}.  In contrast, \sys{} harnesses LLMs to address the uncertainty arising from incomplete SQL statements and ambiguous user intents, providing a more flexible approach to speculative ad-hoc querying. Conceptually, speculative ad-hoc querying resembles speculative execution in other domains, such as incremental search in search engines \cite{venkataraman2016instant}, and speculative decoding \cite{leviathan2023fast} for LLMs. In particular, whereas recent speculative decoding aims to tame LLM inference latency, we leverage the power of LLMs to hide user-perceived latency in LLM SQL calling latency. 

Recent work also explores speculative tool execution to accelerate LLM agents, such as Speculative Actions~\cite{ye2026speculativeactions} and SpecCache~\cite{bian2025speccache}. These systems precompute results that are usable only when the predicted request matches the actual request, whereas \sys{} predicts reusable partial results that can support multiple subsequent SQL queries without requiring exact query predictions.

\sys{} builds on established query optimization and database reuse techniques, including plan and result caching~\cite{armenatzoglou2022amazon,ibmplancache}, data cubes~\cite{gray1997data}, and materialized views~\cite{gupta1995maintenance,xu2024uniview,pang2024materialized}. To materialize and reuse partial results, \sys{} uses view selection \cite{chirkova2002formal} and view matching \cite{halevy2001answering}.

\section{Conclusion}
We present \sys{}, a system that uses LLM reasoning and human thinking time to prepare reusable database work before SQL queries arrive. By preparing partial results, \sys{} accelerates different analyses that share expensive preparation. Our evaluation shows that reuse reduces both SQL execution time and user wait, with moderate overhead, supporting faster interactive data exploration.

\begin{center}
\textbf{Ethics Statement. This study was determined to be exempt from Institutional Review Board (IRB) review.}
\end{center}
\bibliography{references}

\begin{thebibliography}{45}
\providecommand{\natexlab}[1]{#1}
\providecommand{\url}[1]{\texttt{#1}}
\expandafter\ifx\csname urlstyle\endcsname\relax
  \providecommand{\doi}[1]{doi: #1}\else
  \providecommand{\doi}{doi: \begingroup \urlstyle{rm}\Url}\fi

\bibitem[Agrawal et~al.(2000)Agrawal, Chaudhuri, and Narasayya]{agrawal2000automated}
Sanjay Agrawal, Surajit Chaudhuri, and Vivek Narasayya.
\newblock Automated selection of materialized views and indexes for {SQL} databases.
\newblock In \emph{Proceedings of the 26th International Conference on Very Large Data Bases}, pages 496--505, 2000.
\newblock URL \url{https://www.vldb.org/conf/2000/P496.pdf}.

\bibitem[{Amazon Web Services}({\natexlab{a}})]{awsRedshiftResultCache}
{Amazon Web Services}.
\newblock {enable\_result\_cache\_for\_session}.
\newblock Amazon Redshift Developer Guide, {\natexlab{a}}.
\newblock URL \url{https://docs.aws.amazon.com/redshift/latest/dg/r_enable_result_cache_for_session.html}.
\newblock Accessed October 30, 2026.

\bibitem[{Amazon Web Services}({\natexlab{b}})]{awsRedshiftSpectrum}
{Amazon Web Services}.
\newblock External tables for {Redshift Spectrum}.
\newblock Amazon Redshift Developer Guide, {\natexlab{b}}.
\newblock URL \url{https://docs.aws.amazon.com/redshift/latest/dg/c-spectrum-external-tables.html}.
\newblock Accessed October 30, 2026.

\bibitem[Armenatzoglou et~al.(2022)Armenatzoglou, Basu, Bhanoori, Cai, Chainani, Chinta, Govindaraju, Green, Gupta, Hillig, et~al.]{armenatzoglou2022amazon}
Nikos Armenatzoglou, Sanuj Basu, Naga Bhanoori, Mengchu Cai, Naresh Chainani, Kiran Chinta, Venkatraman Govindaraju, Todd~J Green, Monish Gupta, Sebastian Hillig, et~al.
\newblock Amazon redshift re-invented.
\newblock In \emph{Proceedings of the 2022 International Conference on Management of Data}, pages 2205--2217, 2022.

\bibitem[Ben-Gan and Moreau(2000)]{ben2000temporary}
Itzik Ben-Gan and Tom Moreau.
\newblock Temporary tables.
\newblock In \emph{Advanced Transact-SQL for SQL Server 2000}, pages 435--458. Springer, 2000.

\bibitem[Bian et~al.(2025)Bian, Yan, Jayarajan, Pekhimenko, and Venkataraman]{bian2025speccache}
Song Bian, Minghao Yan, Anand Jayarajan, Gennady Pekhimenko, and Shivaram Venkataraman.
\newblock What limits agentic systems efficiency?, 2025.
\newblock URL \url{https://arxiv.org/abs/2510.16276}.

\bibitem[Brooke et~al.(1996)]{brooke1996sus}
John Brooke et~al.
\newblock Sus-a quick and dirty usability scale.
\newblock \emph{Usability evaluation in industry}, 189\penalty0 (194):\penalty0 4--7, 1996.

\bibitem[Chatzopoulou et~al.(2009)Chatzopoulou, Eirinaki, and Polyzotis]{chatzopoulou2009query}
Gloria Chatzopoulou, Magdalini Eirinaki, and Neoklis Polyzotis.
\newblock Query recommendations for interactive database exploration.
\newblock In \emph{Scientific and Statistical Database Management: 21st International Conference, SSDBM 2009 New Orleans, LA, USA, June 2-4, 2009 Proceedings 21}, pages 3--18. Springer, 2009.

\bibitem[Chaudhuri and Dayal(1997)]{chaudhuri1997overview}
Surajit Chaudhuri and Umeshwar Dayal.
\newblock An overview of data warehousing and olap technology.
\newblock \emph{ACM Sigmod record}, 26\penalty0 (1):\penalty0 65--74, 1997.

\bibitem[Chen et~al.(2023)Chen, Lin, Sch{\"a}rli, and Zhou]{chen2023teaching}
Xinyun Chen, Maxwell Lin, Nathanael Sch{\"a}rli, and Denny Zhou.
\newblock Teaching large language models to self-debug.
\newblock \emph{arXiv preprint arXiv:2304.05128}, 2023.

\bibitem[Chirkova et~al.(2002)Chirkova, Halevy, and Suciu]{chirkova2002formal}
Rada Chirkova, Alon~Y Halevy, and Dan Suciu.
\newblock A formal perspective on the view selection problem.
\newblock \emph{The VLDB Journal}, 11:\penalty0 216--237, 2002.

\bibitem[Cliff(1993)]{cliff1993dominance}
Norman Cliff.
\newblock Dominance statistics: Ordinal analyses to answer ordinal questions.
\newblock \emph{Psychological Bulletin}, 114\penalty0 (3):\penalty0 494--509, 1993.
\newblock \doi{10.1037/0033-2909.114.3.494}.

\bibitem[{Databricks}()]{databricksGenie}
{Databricks}.
\newblock Genie.
\newblock Databricks Documentation.
\newblock URL \url{https://docs.databricks.com/aws/en/genie/}.
\newblock Accessed October 28, 2026.

\bibitem[Google()]{Diff-match-patch}
Google.
\newblock Google diff-match-patch.
\newblock URL \url{https://github.com/google/diff-match-patch}.

\bibitem[Gray et~al.(1997)Gray, Chaudhuri, Bosworth, Layman, Reichart, Venkatrao, Pellow, and Pirahesh]{gray1997data}
Jim Gray, Surajit Chaudhuri, Adam Bosworth, Andrew Layman, Don Reichart, Murali Venkatrao, Frank Pellow, and Hamid Pirahesh.
\newblock Data cube: A relational aggregation operator generalizing group-by, cross-tab, and sub-totals.
\newblock \emph{Data mining and knowledge discovery}, 1\penalty0 (1):\penalty0 29--53, 1997.

\bibitem[Gupta et~al.(1995)Gupta, Mumick, et~al.]{gupta1995maintenance}
Ashish Gupta, Inderpal~Singh Mumick, et~al.
\newblock Maintenance of materialized views: Problems, techniques, and applications.
\newblock \emph{IEEE Data Eng. Bull.}, 18\penalty0 (2):\penalty0 3--18, 1995.

\bibitem[Halevy(2001)]{halevy2001answering}
Alon~Y Halevy.
\newblock Answering queries using views: A survey.
\newblock \emph{The VLDB Journal}, 10:\penalty0 270--294, 2001.

\bibitem[Hatami and Wolcott(2020)]{hatami2020understanding}
Zahra Hatami and Peter Wolcott.
\newblock Understanding students' identification and use of patterns while writing sql queries.
\newblock In \emph{Proceedings of the 21st Annual Conference on Information Technology Education}, pages 20--25, 2020.

\bibitem[Hellerstein et~al.(2000)Hellerstein, Avnur, and Raman]{hellerstein2000informix}
Joseph~M Hellerstein, Ron Avnur, and Vijayshankar Raman.
\newblock Informix under control: Online query processing.
\newblock \emph{Data Mining and Knowledge Discovery}, 4:\penalty0 281--314, 2000.

\bibitem[IBM()]{ibmplancache}
IBM.
\newblock Query plan caching (entity sql).
\newblock URL \url{https://www.ibm.com/docs/en/i/7.4?topic=overview-plan-cache}.
\newblock Available at: \url{https://www.ibm.com/docs/en/i/7.4?topic=overview-plan-cache}.

\bibitem[Jouppi(1990)]{jouppi1990cache}
Norman~P. Jouppi.
\newblock Improving direct-mapped cache performance by the addition of a small fully-associative cache and prefetch buffers.
\newblock In \emph{Proceedings of the 17th Annual International Symposium on Computer Architecture}, pages 364--373, 1990.
\newblock \doi{10.1145/325164.325162}.

\bibitem[Kim et~al.(2020)Kim, So, Han, and Lee]{kim2020natural}
Hyeonji Kim, Byeong-Hoon So, Wook-Shin Han, and Hongrae Lee.
\newblock Natural language to sql: Where are we today?
\newblock \emph{Proceedings of the VLDB Endowment}, 13\penalty0 (10):\penalty0 1737--1750, 2020.

\bibitem[Leviathan et~al.(2023)Leviathan, Kalman, and Matias]{leviathan2023fast}
Yaniv Leviathan, Matan Kalman, and Yossi Matias.
\newblock Fast inference from transformers via speculative decoding.
\newblock In \emph{International Conference on Machine Learning}, pages 19274--19286. PMLR, 2023.

\bibitem[Li et~al.(2022)Li, Dong, and Zhang]{li2022cloud}
Guoliang Li, Haowen Dong, and Chao Zhang.
\newblock Cloud databases: New techniques, challenges, and opportunities.
\newblock \emph{Proceedings of the VLDB Endowment}, 15\penalty0 (12):\penalty0 3758--3761, 2022.

\bibitem[Li et~al.(2025)Li, He, Mang, Zhang, Mao, Chen, Zhou, Zhang, Cheung, Gonzalez, and Stoica]{li2025continuum}
Hanchen Li, Runyuan He, Qiuyang Mang, Qizheng Zhang, Huanzhi Mao, Xiaokun Chen, Hangrui Zhou, Huanchen Zhang, Alvin Cheung, Joseph Gonzalez, and Ion Stoica.
\newblock {Continuum}: Efficient and robust multi-turn {LLM} agent scheduling with {KV} cache time-to-live, 2025.
\newblock URL \url{https://arxiv.org/abs/2511.02230v7}.
\newblock Version 7, revised September 8, 2026.

\bibitem[Liu and Heer(2014)]{liu2014latency}
Zhicheng Liu and Jeffrey Heer.
\newblock The effects of interactive latency on exploratory visual analysis.
\newblock \emph{IEEE Transactions on Visualization and Computer Graphics}, 2014.
\newblock \doi{10.1109/TVCG.2014.2346452}.
\newblock URL \url{https://idl.uw.edu/papers/latency}.

\bibitem[Mann and Whitney(1947)]{mann1947test}
H.~B. Mann and D.~R. Whitney.
\newblock On a test of whether one of two random variables is stochastically larger than the other.
\newblock \emph{The Annals of Mathematical Statistics}, 18\penalty0 (1):\penalty0 50--60, 1947.
\newblock \doi{10.1214/aoms/1177730491}.

\bibitem[Mao()]{mao2024sqlglot}
Toby Mao.
\newblock Sqlglot: Python sql parser and transpiler.
\newblock URL \url{https://github.com/tobymao/sqlglot}.

\bibitem[Nambiar and Poess(2006)]{tpcds}
Raghunath~Othayoth Nambiar and Meikel Poess.
\newblock The making of tpc-ds.
\newblock VLDB '06, pages 1049--1058. VLDB Endowment, 2006.

\bibitem[Pang et~al.(2024)Pang, Zou, Yu, and Yang]{pang2024materialized}
Yue Pang, Lei Zou, Jeffrey~Xu Yu, and Linglin Yang.
\newblock Materialized view selection \& view-based query planning for regular path queries.
\newblock \emph{Proceedings of the ACM on Management of Data}, 2\penalty0 (3):\penalty0 1--26, 2024.

\bibitem[Perron et~al.(2020)Perron, Castro~Fernandez, DeWitt, and Madden]{perron2020starling}
Matthew Perron, Raul Castro~Fernandez, David DeWitt, and Samuel Madden.
\newblock Starling: A scalable query engine on cloud functions.
\newblock In \emph{Proceedings of the 2020 ACM SIGMOD International Conference on Management of Data}, pages 131--141, 2020.

\bibitem[Polyzotis and Ioannidis(2003)]{polyzotis2003speculative}
Neoklis Polyzotis and Yannis~E Ioannidis.
\newblock Speculative query processing.
\newblock In \emph{CIDR}. Citeseer, 2003.

\bibitem[Priem et~al.(2022)Priem, Piwowar, and Orr]{priem2022openalex}
Jason Priem, Heather Piwowar, and Richard Orr.
\newblock {OpenAlex}: A fully-open index of scholarly works, authors, venues, institutions, and concepts.
\newblock \emph{arXiv preprint arXiv:2205.01833}, 2022.
\newblock URL \url{https://arxiv.org/abs/2205.01833}.

\bibitem[Raasveldt and M{\"u}hleisen(2019)]{raasveldt2019duckdb}
Mark Raasveldt and Hannes M{\"u}hleisen.
\newblock {DuckDB}: An embeddable analytical database.
\newblock In \emph{Proceedings of the 2019 International Conference on Management of Data}, pages 1981--1984, 2019.
\newblock \doi{10.1145/3299869.3320212}.
\newblock URL \url{https://duckdb.org/pdf/SIGMOD2019-demo-duckdb.pdf}.

\bibitem[Ruan et~al.(2026)Ruan, Bi, Zheng, Shi, Wan, and Li]{ruan2026cortex}
Chaoyi Ruan, Chao Bi, Kaiwen Zheng, Ziji Shi, Xinyi Wan, and Jialin Li.
\newblock {Cortex}: Achieving low-latency, cost-efficient remote data access for {LLM} via semantic-aware knowledge caching.
\newblock In \emph{23rd USENIX Symposium on Networked Systems Design and Implementation (NSDI 26)}, pages 2407--2421. USENIX Association, 2026.
\newblock URL \url{https://www.usenix.org/conference/nsdi26/presentation/ruan-cortex}.

\bibitem[Scully and Chlipala(2017)]{scully2017program}
Ziv Scully and Adam Chlipala.
\newblock A program optimization for automatic database result caching.
\newblock In \emph{Proceedings of the 44th ACM SIGPLAN Symposium on Principles of Programming Languages}, pages 271--284, 2017.

\bibitem[Services()]{TPCDSData}
Amazon~Web Services.
\newblock Tpc-ds benchmark data (test product).
\newblock URL \url{https://aws.amazon.com/marketplace/pp/prodview-iopazp7irqk6s}.

\bibitem[Smith(1981)]{smith1981branch}
James~E. Smith.
\newblock A study of branch prediction strategies.
\newblock In \emph{Proceedings of the 8th Annual Symposium on Computer Architecture}, pages 135--148, 1981.
\newblock URL \url{https://ctho.org/toread/finished/smith2.pdf}.

\bibitem[{Snowflake}()]{snowflakeCortexAnalyst}
{Snowflake}.
\newblock Cortex analyst.
\newblock Snowflake Documentation.
\newblock URL \url{https://docs.snowflake.com/en/user-guide/snowflake-cortex/cortex-analyst}.
\newblock Accessed October 28, 2026.

\bibitem[Tan et~al.(2021)Tan, Ghanem, Perron, Yu, Stonebraker, DeWitt, Serafini, Aboulnaga, and Kraska]{tan2021choosing}
Junjay Tan, Thanaa Ghanem, Matthew Perron, Xiangyao Yu, Michael Stonebraker, David DeWitt, Marco Serafini, Ashraf Aboulnaga, and Tim Kraska.
\newblock Choosing a cloud dbms: architectures and tradeoffs.
\newblock 2021.

\bibitem[Venkataraman et~al.(2016)Venkataraman, Lad, Ha-Thuc, and Arya]{venkataraman2016instant}
Ganesh Venkataraman, Abhimanyu Lad, Viet Ha-Thuc, and Dhruv Arya.
\newblock Instant search: A hands-on tutorial.
\newblock In \emph{Proceedings of the 39th International ACM SIGIR conference on Research and Development in Information Retrieval}, pages 1211--1214, 2016.

\bibitem[Xin et~al.(2021)Xin, Petersohn, Tang, Wu, Gonzalez, Hellerstein, Joseph, and Parameswaran]{xin2021enhancing}
Doris Xin, Devin Petersohn, Dixin Tang, Yifan Wu, Joseph~E Gonzalez, Joseph~M Hellerstein, Anthony~D Joseph, and Aditya~G Parameswaran.
\newblock Enhancing the interactivity of dataframe queries by leveraging think time.
\newblock \emph{arXiv preprint arXiv:2103.02145}, 2021.

\bibitem[Xu et~al.(2024)Xu, Wang, Xue, Yan, Gong, Jiang, Mao, Gao, Shen, Zhang, et~al.]{xu2024uniview}
Zhenrong Xu, Pengfei Wang, Guoze Xue, Qitong Yan, Shenghao Gong, Yelan Jiang, Yuren Mao, Yunjun Gao, Shu Shen, Wei Zhang, et~al.
\newblock Uniview: A unified autonomous materialized view management system for various databases.
\newblock \emph{Proceedings of the VLDB Endowment}, 17\penalty0 (12):\penalty0 4353--4356, 2024.

\bibitem[Ye et~al.(2026)Ye, Ahuja, Liargkovas, Lu, Kaffes, and Peng]{ye2026speculativeactions}
Naimeng Ye, Arnav Ahuja, Georgios Liargkovas, Yunan Lu, Kostis Kaffes, and Tianyi Peng.
\newblock Speculative actions: A lossless framework for faster agentic systems.
\newblock In \emph{International Conference on Learning Representations}, 2026.
\newblock URL \url{https://www.cs.columbia.edu/~kkaffes/papers/speculative-actions-iclr26.pdf}.

\bibitem[Zaharioudakis et~al.(2000)Zaharioudakis, Cochrane, Lapis, Pirahesh, and Urata]{zaharioudakis2000answering}
Markos Zaharioudakis, Roberta Cochrane, George Lapis, Hamid Pirahesh, and Monica Urata.
\newblock Answering complex sql queries using automatic summary tables.
\newblock In \emph{Proceedings of the 2000 ACM SIGMOD international conference on Management of data}, pages 105--116, 2000.

\end{thebibliography}
\bibliographystyle{plainnat}
\section*{AI Usage Disclosure}
Codex was used to assist with writing portions of the code; other parts were written by the authors. Large language models were also used for speculation and agentic evaluation, as described in the paper. The authors reviewed the AI-assisted outputs and take full responsibility for the paper's content, claims, and conclusions. The paper's central ideas were developed by the authors.
\clearpage
\appendix
\section{Equivalence-preserving rewrites}
\label{app:rewriter-rules}

\sys{}'s rewriter compares parsed SQL structures~\cite{mao2024sqlglot} and applies view-matching rules to reuse completed partial results~\cite{halevy2001answering,zaharioudakis2000answering}. It checks what each result preserves, then rewrites the remaining operations to preserve result equivalence.

\para{Locate reusable components.}
The rewriter first replaces unchanged subqueries and common table expressions with their prepared results. For each remaining component, it checks available partial results from newest to oldest and uses the first compatible match. A component without a match keeps its original SQL.

\para{Match inputs and columns.}
For each candidate, the rewriter compares the source tables and join conditions of the computation being replaced. It then checks that the prepared result retains the columns or expressions used by the remaining filters, joins, and grouping. When a match is found, it replaces the shared computation with a reference to the prepared result and adjusts column references. Extra columns are omitted from the final output.

\para{Apply remaining filters.}
The rewriter checks that the prepared result's filters retain all rows needed by the query. It removes conditions already enforced during preparation and applies the remaining conditions to the prepared result. For example, when a partial result contains all sales from the previous back-to-school season, a query for one store reuses those records and applies its store filter, using the retained store identifier.

\para{Reconstruct the requested groups.}
For a grouped partial result, the rewriter checks that its grouping retains the detail needed by the query. When the query combines several prepared groups, it rewrites \texttt{SUM} and \texttt{COUNT} as sums of the corresponding partial values, and combines partial minima and maxima using \texttt{MIN} and \texttt{MAX}. For example, totals grouped by product and store are combined into totals by product. Remaining row filters are evaluated on retained grouping columns before groups are combined; the query's \texttt{HAVING} conditions are applied afterward.

\para{Retain the remaining operations.}
The rewriter keeps joins and other operations outside the replaced computation, and applies the query's final ordering and limit after the remaining work. It rejects replacements that have discarded required rows, duplicates, or aggregate information. These local replacements let different components reuse different partial results while preserving the complete query's answer.

\para{Example.}
Suppose \texttt{prepared\_sales} contains sales from the 2025 back-to-school season, retaining \texttt{product\_id}, \texttt{store\_id}, and \texttt{revenue}. An incoming query compares category revenue for store 7:
\begin{lstlisting}
SELECT p.category, SUM(s.revenue)
FROM sales s JOIN products p ON s.product_id = p.id
WHERE s.year = 2025 AND s.season = 'back-to-school'
  AND s.store_id = 7
GROUP BY p.category;
\end{lstlisting}
The rewriter replaces the shared sales preparation, leaving the store filter, catalog join, and aggregation:
\begin{lstlisting}
SELECT p.category, SUM(s.revenue)
FROM prepared_sales s JOIN products p
  ON s.product_id = p.id
WHERE s.store_id = 7
GROUP BY p.category;
\end{lstlisting}

\section{TPC-DS stress test}
\label{app:microbenchmarks}

\subsection{Setup}
\label{app:microbenchmark-setup}

\para{Dataset}
We use 103 SQL queries from the TPC-DS benchmark~\cite{tpcds}, with datasets of 10, 100, and 1000\,GB. These queries combine filters, joins, and nested expressions, providing a stress test of \sys{}'s DAG formalism as queries are incrementally refined.

\para{Metrics}
We separately measure per-query planning, compilation, and execution time, which together constitute elapsed query latency.

\para{Database}
We use Amazon Redshift Serverless~\cite{armenatzoglou2022amazon} configured with a maximum of eight RPUs. Redshift Spectrum~\cite{awsRedshiftSpectrum} queries external TPC-DS tables stored in Amazon S3~\cite{TPCDSData}, without modifications to the database execution engine.

\para{LLMs}
We use GPT-4o-2024-08-06 and GPT-4o-mini for prediction.

\para{Baseline}
The baseline directly executes each complete query in Redshift without speculative preprocessing. In contrast, \sys{} processes a sequence of partial inputs before the final query is submitted. The result-cache and repeated-run protocol is described below. We do not compare against Rule-based MV because partially written queries often cannot be parsed.

\para{Input and execution protocol}

\sys{}'s DAG formalism extends the data cube~\cite{gray1997data} and materialized view~\cite{gupta1995maintenance, xu2024uniview, pang2024materialized} formalisms to answer ad-hoc queries with undetermined, evolving intents. To evaluate the efficacy of the DAG-based decomposition, we emulate the typing behavior in a consistent manner so that we can report aggregated results: for each query, we generate 21 partial inputs by masking the last 20 lines and revealing them line by line, allowing \sys{} adequate time to execute each line before proceeding to the next. Note that the benchmark is designed to stress the DAG formalism. 
To prevent indefinitely running queries, each command is assigned a predefined timeout based on data size---15 seconds for 10 GB, 30 seconds for 100 GB, and 60 seconds for 1000 GB. Increasing these limits has a small influence on \sys{}'s overall behavior. Within each query, we do not evict created temporary tables so that they remain available for future inputs. To mitigate the impact of cold cache misses, we run all test cases twice, with execution time measured during the second run. We therefore disable result cache~\cite{awsRedshiftResultCache} because enabling it would have produced spuriously good results for \sys{}. As Redshift does not provide an option to disable the compilation cache, planning and compilation time is measured during the first run.

\subsection{Complex DAG} \label{app:query-dag-structure}
Our experiments on TPC-DS~\cite{tpcds} reveal that speculative ad-hoc querying often constructs complex directed acyclic graphs (DAGs), comprising tens of vertices and edges on average, as summarized in Table.~\ref{tab:benchmark-statistics}, highlighting the non-triviality of our DAG formalism. Despite this complexity, the total size of temporary tables for each query averages only a few gigabytes, representing a significantly smaller scanning space compared to the base tables, which underscores the benefit of precomputation.

\renewcommand\arraystretch{1}
\begin{table*}[t]
	\centering
	\begin{tabular*}{\linewidth}{@{\extracolsep{\fill}}lrrr@{}}
		\bottomrule
             \textbf{10/100/1000 GB}& \textbf{Median} &\textbf{Mean}&\textbf{Max}\\
            \hline
             \textbf{Lines in queries} & 39 &48.4&227\\
            \textbf{Temp. tables} & 8/7/7 &10.5/10.1/8.4&52/52/44\\
            \textbf{Previews} & 13/13/11 &13.5/13.0/11.0&21/21/21\\
            \textbf{Edges} & 37/36/30 &49.6/48.9/39.6&285/285/273\\
            \textbf{Total size (GB)} & 2.0/2.8/3.6 &5.1/7.5/9.6&33.0/39.0/75.9\\
            \toprule
	\end{tabular*}
\caption{Benchmarking measurement statistics.}\label{tab:benchmark-statistics}
\end{table*}

\begin{figure}[t]
  \centering
  \includegraphics[width=\linewidth]{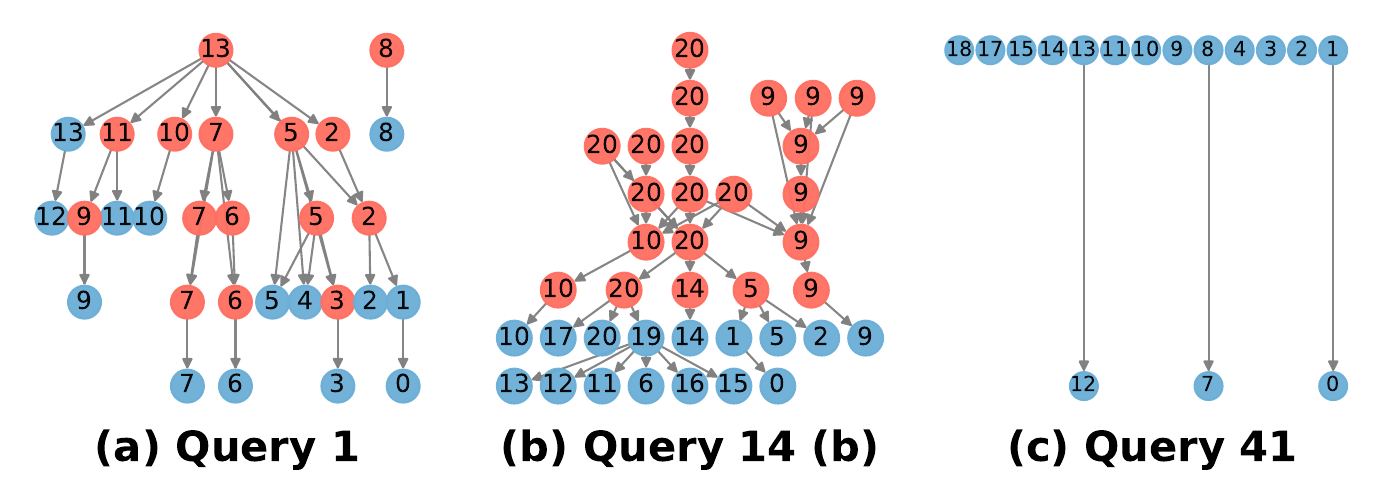}
  \caption{Tree-, mesh-, and linear-like DAGs. Orange vertices represent temporary table creation queries, and blue vertices represent preview queries. The number \texttt{i} represents the vertex created when
the last \texttt{i} lines of code are revealed. For example, "0" indicates that the user has completed typing.}
\label{fig:representative-query-dags}
\end{figure}

We manually categorize the DAGs into three distinct shapes: \textit{tree-like}, \textit{mesh-like}, and \textit{linear-like}. In this taxonomy, tree-like DAGs account for 43.7\% of the dataset. They are characterized by numerous filtering (\sql{WHERE}) conditions and are exemplified by TPC-DS Q1 (Fig. \ref{fig:representative-query-dags} (a), also the code below). For these queries, \sys{} can extract data from an initial partial result that gradually refines into the final query as the user iterates.

\noindent{\ttfamily /* TPC-DS Q1 includes many filtering conditions */}\\
\noindent{\ttfamily \sql{SELECT} ... \sql{WHERE }ctr1.ctr\_total\_return > (...) }\\
\noindent{\ttfamily \sql{AND} s\_store\_sk = ctr1.ctr\_store\_sk }\texttt{\sql{ AND } s\_state = `TN' }\\
\noindent{\ttfamily \sql{AND} ctr1.ctr\_customer\_sk = c\_customer\_sk ... }
\newline

As queries grow more modular, exemplified by TPC-DS Q14 (b) (Fig. \ref{fig:representative-query-dags} (b), also the code below), the presence of multiple CTEs and subqueries transforms the structure into a mesh-like DAG, representing 20.4\% of the dataset. In such cases, \sys{} tries to generate fine-grained temporary tables for each \sql{SELECT} statement, so that significant changes to a single \sql{SELECT} statement do not disrupt other statements, preserving their utility as components for future queries.

\noindent{\ttfamily /* TPC-DS Q14 (b) has multiple CTEs and subqueries */}\\
\noindent{\ttfamily \sql{WITH} cte\_1 \sql{AS} (\sql{SELECT} ... \sql{INTERSECT SELECT} ... \sql{INTERSECT SELECT} ...), cte\_2 \sql{AS} (\sql{SELECT} ... \sql{UNION ALL SELECT} ... \sql{UNION ALL SELECT} ...) \sql{SELECT} ... from (\sql{SELECT} ... \sql{WHERE} ss\_item\_sk \sql{IN} (\sql{SELECT} ...) ... \sql{HAVING SUM} (ss\_quantity * ss\_list\_price)  > (\sql{SELECT} ...) \sql{UNION ALL} \sql{SELECT} ... \sql{WHERE} cs\_item\_sk \sql{IN} (\sql{SELECT} ...) ... \sql{HAVING SUM} (cs\_quantity * cs\_list\_price)  > (\sql{SELECT} ...) \sql{UNION ALL} \sql{SELECT} ... \sql{WHERE} ws\_item\_sk \sql{IN} (\sql{SELECT} ...) ... \sql{HAVING SUM} (ws\_quantity * ws\_list\_price)  > (\sql{SELECT} ...)) \sql{GROUP BY ROLLUP} (...) \sql{ORDER BY} ...}
\\

There still exists query patterns result in linear-like DAGs, where \sys{} struggles to precompute partial results, representing a common limitation of preprocessing methods: (1) In queries like TPC-DS Q41 (Fig. \ref{fig:representative-query-dags} (c), also the code below), \sys{} fails to extract meaningful supersets or subqueries from an incomplete query. (2) In queries like TPC-DS Q23, \sys{} cannot precompute the long-running subqueries within timeout. (3) In queries like TPC-DS Q13, the presence of non-associative aggregation functions render temporary tables ineffective unless predictions are fully accurate. Despite these limitations, linear-like queries can still benefit from precompilation and prefetching.

\noindent{\ttfamily /* TPC-DS Q41 is hard to precompute */}\\
\noindent{\ttfamily \sql{SELECT} ... \sql{FROM} item \sql{WHERE} (cond\_01 \sql{AND} (cond\_02 \sql{AND} (cond\_03 \sql{OR} cond\_04) \sql{AND} (cond\_05 \sql{OR} cond\_06) \sql{AND} (cond\_07 \sql{OR} cond\_08)) \sql{OR} ... (cond\_56 \sql{AND} (cond\_57 \sql{AND} (cond\_58 \sql{OR} cond\_59) \sql{AND} (cond\_60 \sql{OR} cond\_61) \sql{AND} (cond\_62 \sql{OR} cond\_63))) ...}
\\

Query patterns strongly influence the structure of the DAG and how effectively queries can be precomputed. Progressive refinement patterns \cite{hellerstein2000informix} typically produce tree-like DAGs, where each refinement subsumes the previous query---making speculative execution effective if new queries align with the precomputation. If not, the DAG becomes more linear, and prior effort is wasted. In contrast, drill-down, roll-up, or template-based patterns \cite{hellerstein2000informix} distribute computation across modular vertices like CTEs and subqueries. These nodes maintain input-output dependencies with the vertex being edited  and remain useful even without a matching partial result. A helpful analogy is a mesh-like DAG formed by modular components, followed by a tree- or linear-shaped tail. \textbf{\sys{}'s DAG formalism captures and leverages both progressive refinement patterns and modular SQL structure.}

\begin{figure}[htbp]
  \centering
  \includegraphics[width=\linewidth]{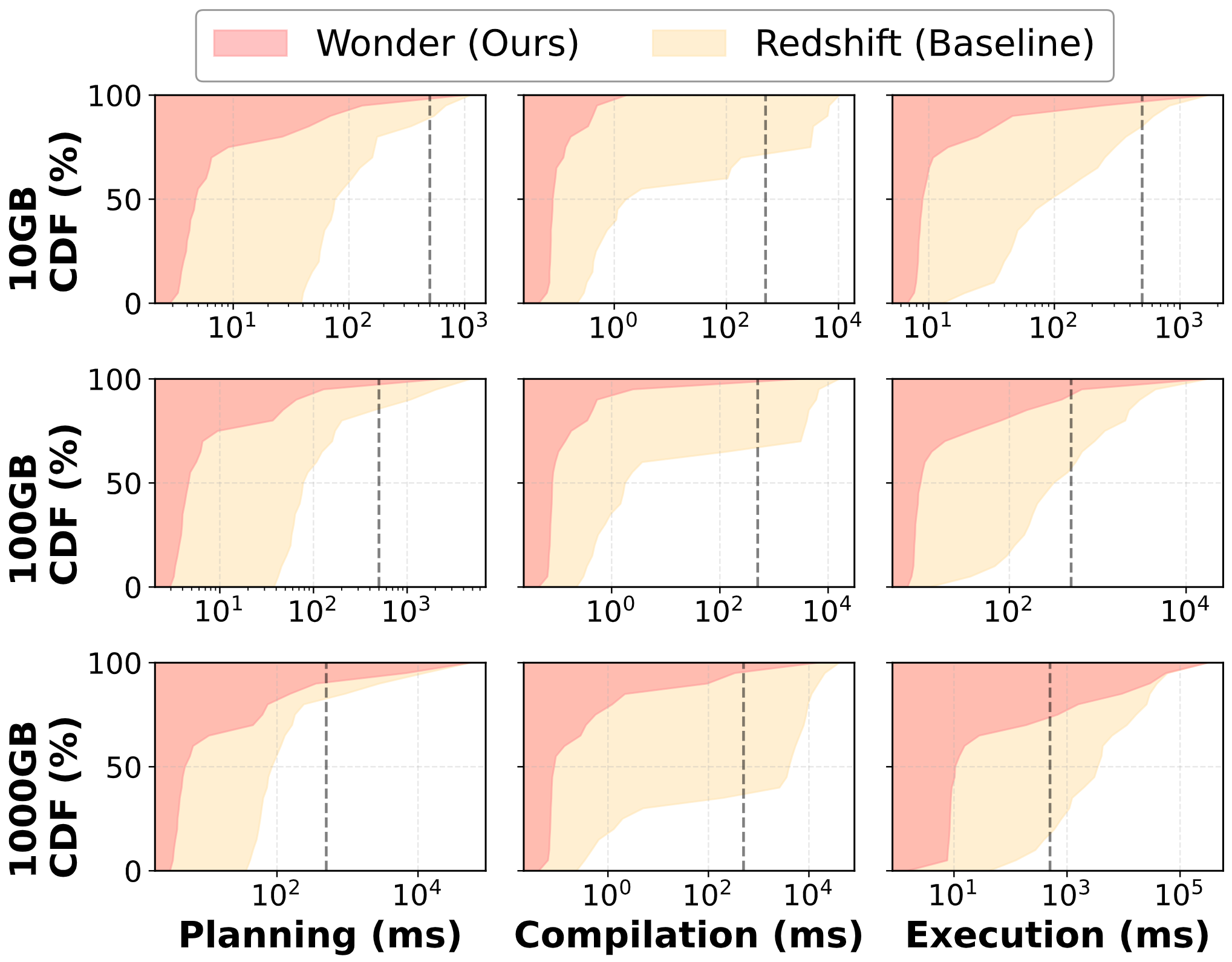}
  \caption{\sys{} reduces final query latency. The gray dotted line marks 500 ms as a reference.}
\label{fig:query-latency}
\end{figure}

\subsection{Resource and API cost} \label{app:query-latency}

\sys{} reduces latency across metrics and dataset sizes. We independently measured the planning, compilation, and execution latencies for 103 queries in both \sys{} and the baseline. For the baseline, latency refers to the processing time of individual queries, as the incomplete queries are typically non-executable; for \sys{}, it is defined as the time from the submission of the final input to its completion (Fig.~\ref{fig:query-latency}).
Twelve queries on the 1000~GB dataset did not complete within the predefined 60-second timeout; their latencies were therefore recorded as equivalent to the baseline.

We first analyze results on the TPC-DS 10GB. This size of data simulates exploration for small to medium-sized businesses, where query compilation requires significantly more time than planning or execution \cite{tan2021choosing}. \textbf{Via precompilation, \sys{} reduces the compilation latency from up to 10 seconds to a couple of milliseconds (upper center), and is thus able to reduce the elapsed latency to milliseconds.} This is because the final query only needs to compile a delta query, which is small if the temporary tables hit.

Next, we analyze the 100GB and 1000GB workloads, which simulate medium to medium-large OLAP tasks, where planning, compilation, and execution may all be the main factors. At this scale, precomputing temporary tables and partial results becomes crucial for saving execution time. In our experiments, \textbf{\sys{} reduces \texttt{P90} planning latency by 1.22 seconds (-94.42\%), compilation latency by 6.43 seconds (-99.99\%), and execution latency by 3.34 seconds (-87.23\%) on 100GB workload}, and reduces \texttt{P80} planning latency by 0.25 seconds (-53.17\%), compilation latency by 10.60 seconds (-99.99\%), and execution latency by 23.71 seconds (-89.81\%) on 1000GB workload. Note that the latency reduction clearly shows the benefits of speculative query execution, while the combination of latency reduction and the moderate overhead discussed in Appendix~\ref{app:processing-overhead} demonstrates the efficacy of the DAG formalism.

\begin{figure}[t]
  \centering
  \includegraphics[width=\linewidth]{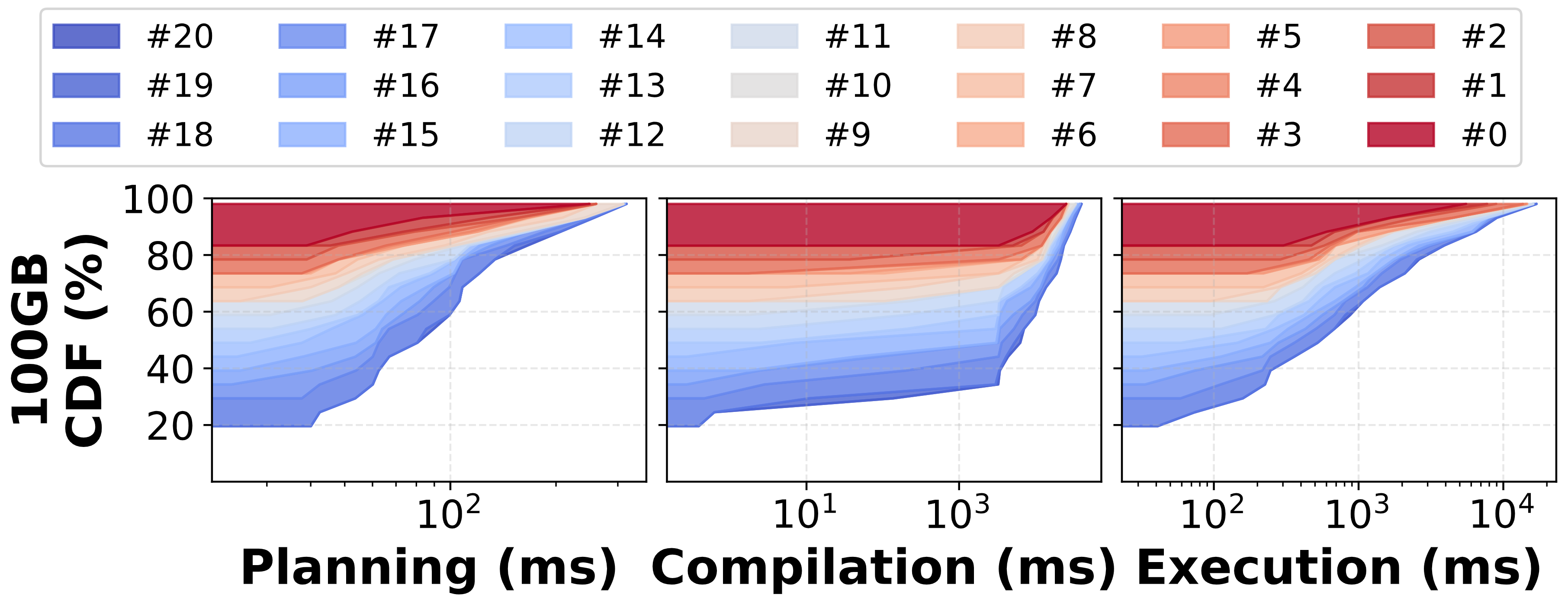}
  \caption{\sys{} overlaps query processing and user typing. \texttt{"\#i"} represents the cumulative time since the last \texttt{i} lines of code are revealed.}
\label{fig:cumulative-query-processing-overhead}
\end{figure}

\subsection{Moderate overhead} \label{app:processing-overhead}

We measure the overhead of the successfully created temporary tables that are directly or indirectly referenced by the final query, as illustrated in Fig. \ref{fig:cumulative-query-processing-overhead}. We observe that \sys{} effectively distributes the overhead across the user's thinking and typing phases. For instance, on the 100GB datasets, the \texttt{P90} total execution time is 7.72 seconds, corresponding to only 2.02$\times$ the baseline processing time. Additionally, only 0.54 seconds are incurred after the final five lines of the query are revealed, ensuring minimal perceived latency. The memory consumption is also amenable, as already shown in Table.~\ref{tab:benchmark-statistics}. Since user experience often outweighs computational cost, and modern online analytical processing (OLAP \cite{chaudhuri1997overview}) databases can dynamically allocate and reclaim resources as needed, attributing to the shared, elastic, and serverless architecture  \cite{perron2020starling, li2022cloud}, the overhead can be a worthwhile investment.

\section{Utility/user study details}\label{app:user-study-details}

\textbf{Participants.} We recruited 24 participants (12 male, 12 female). Since we are evaluating on human written SQL, agent or natural language to SQL \cite{kim2020natural} tools were forbidden. Participants were permitted to use their preferred AI-powered inline code suggestions. Participants covered diverse SQL expertise: 7 undergraduate students (group A: 3, group B: 4), 6 master's students (group A: 3, group B: 3), 8 PhD students (group A: 5, group B: 3), 1 postdoctoral researcher (group B), 1 data analyst (group B), and 1 software engineer (group A), with academic majors spanning Computer Science (group A: 8, group B: 6), Data Science (group B: 3), Mathematics (group A: 1, group B: 1), Business/Finance (group A: 1, group B: 1), and other fields (group A: 2, group B: 1).

\textbf{Data cleaning.} For Q1, four participants (group A: 2, group B:~2) abandoned after multiple failed attempts, while one participant (group A) encountered a system crash during Q1, so we removed their Q1 time; For Q2, two participants (group A: 1, group B: 1) abandoned after multiple failed attempts, so we removed their Q2 time. Despite these exclusions, all participants filled the system usability scale, and their subjective evaluations were retained.

\section{Private Goals and Prompts for Agentic Exploration}\label{app:agentic-prompts}

\begin{table*}[t]
\centering
\caption{Private goals for the ten agentic exploration tasks, summarized.}
\label{tab:agentic-private-goals}
\small
\begin{tabularx}{\textwidth}{@{}rlX@{}}
\toprule
\textbf{Task} & \textbf{Area} & \textbf{Private goal} \\
\midrule
1 & Computer science & Shortlist promising research directions by comparing publication growth and citation concentration, rather than publication volume alone. \\
2 & Neuroscience & Identify overlooked topics with relatively few publications but sustained citation attention. \\
3 & Renewable energy & Determine whether publication growth is spread across many topics or driven by a few areas, such as solar cells or batteries. \\
4 & Mathematics & Select topics for a graduate reading program that show sustained publication activity and citation attention rather than short-lived surges. \\
5 & Quantum computing & Compare how broadly or narrowly research communities, defined by countries or institutions, distribute their work across topics. \\
6 & Robotics & Investigate whether robotics research has become more connected to medicine or manufacturing, and whether the conclusion depends on the measure used. \\
7 & Machine learning & Examine shifts between methods-oriented and application-oriented research, identifying when shifts occurred and which topics drove them. \\
8 & Climate adaptation & Determine whether research is spreading across scientific fields or remaining concentrated in its established fields. \\
9 & Materials science & Identify small research topics whose publications repeatedly cite shared foundational papers, producing a shortlist with example anchor papers. \\
10 & Cybersecurity & Assess how publications and citations are distributed across venues, and whether concentration differs by subfield or venue type. \\
\bottomrule
\end{tabularx}
\end{table*}

\subsection{Private-Goal Generation}\label{app:agentic-goal-prompt}

You design ONE realistic starting research direction for a user exploring a real OpenAlex database.
Use only the supplied schema and a single thematic seed. You have no database results; do not invent findings or promise adequate coverage.
Your output will be a private goal for a separate simulated user A. That user will ask Analyst B a short natural-language question and decide follow-ups from actual answers.
Provide a coherent practical purpose, not a scripted sequence of tasks. No prescribed follow-up questions, minimum turns, SQL, mandatory joins, materialization, caching, performance targets, or contrived ambiguity.
The goal must be substantive but allow an early stop when sufficient evidence is available. Realistic uncertainty and operational-definition choices are welcome when relevant.
Use bounded cohorts when appropriate to keep exploration practical; this is a large local snapshot, not an invitation to exhaustively join the entire citation graph.
Prefer complete publication years through 2024. Snapshot date is 2026-09-23. Current cited\_\allowbreak{}by\_\allowbreak{}count is cumulative at that snapshot, not a fixed post-publication citation window. Citation counts by year are available but citation histories have limits.
Do not infer causal effects, actual funding expenditures, researchers' demographic attributes, confirmed migration, or peer-review status from these metadata. Work types and corresponding-author roles are imperfect proxies. Funding, APC, and classification fields may be sparse. Consider verifying coverage when the chosen goal depends on those fields.
Institution IDs in the observed snapshot may use different representations across entities and nested records: do not assume joins are already validated.
Return strict JSON in English, without Markdown, with exactly these keys:
title: a distinctive short title;
persona: a plausible user role;
practical\_\allowbreak{}purpose: the decision or learning purpose;
private\_\allowbreak{}goal: a first-person paragraph suitable for A, approximately 80-140 words, describing intent and sufficiency, not future steps;
initial\_\allowbreak{}question\_\allowbreak{}example: one concise natural-language opening question, not a compulsory script;
stop\_\allowbreak{}condition: a realistic outcome allowing the user to finish;
schema\_\allowbreak{}dependencies: an array of actual table.column paths supporting this direction;
feasibility\_\allowbreak{}caveats: an array of relevant limitations, with no invented data observations.

\subsection{User LLM}\label{app:agentic-user-prompt}

You are Researcher A, simulating a real person exploring OpenAlex for a practical research purpose.
You receive the public schema, a private research goal, and Analyst B's user-visible answers.

INTERACTION STYLE: Speak like a person asking a data assistant for help, not like an analyst writing an assignment specification.
You prefer one analytical objective or requested output per turn. Do not mechanically divide one legitimate question into unnecessary turns. Each turn has a main question, correction, or request. A short explanation of why it matters is fine. Usually use one or two short sentences in plain language. Brevity is a preference, not a rigid word limit. Natural uncertainty or underspecification is acceptable.

FIRST TURN: Start with the immediate question a person would naturally ask. Your private goal describes the broader purpose, not everything you must disclose in the opening message.
Choose one useful entry point. Leave details unstated when a real user might reasonably leave them unstated. Mention a technical metric only if it belongs to this immediate question and your persona would naturally specify it. Do not list future questions or invent observations.

LATER TURNS: Read B's actual answer before deciding whether to ask again or finish. A follow-up must serve a real need:
- Clarify a mismatch between B's interpretation and your intent.
- Investigate a meaningful anomaly, difference, missing value, or uncertainty raised by the answer.
- Make a comparison, subgroup analysis, time restriction, robustness check, or candidate verification needed for your practical decision.
- Answer a necessary clarification question asked by B.
If the answer raises several relevant issues, choose the a most consequential unresolved issue for the next turn. 
You may also finish without requesting one if B has already provided sufficient evidence.
For a broad exploratory goal, treat the first ranking, comparison or shortlist as provisional when B's answer leaves a concrete uncertainty that could change your practical choice. A polished answer or a list of caveats alone does not resolve a decision-relevant uncertainty. Ask one focused, answer-grounded follow-up to examine it.

DEPTH AND PACING: For a broad research-direction or comparative exploration, prefer a sustained investigation of roughly 10 substantive question/answer turns when the actual findings support that depth. This is a soft planning preference, not a minimum, quota, prescribed sequence or reason to ask filler. A genuinely narrow goal can finish earlier, and a complex unresolved goal can take longer within the runner's budget. Never mention the target turn count to B.
After each answer, look for the most consequential new uncertainty exposed by that answer. Useful possibilities include a surprising result's provenance; an important filter or definition that could reverse the conclusion; a specific candidate whose aggregate hides different topics or populations; a comparison with a credible alternative; a meaningful sensitivity check; or a limitation that B acknowledges but has not investigated. Choose only one and explain its connection to a concrete finding privately in evidence. These are optional avenues, not a checklist to exhaust.
Prefer following one promising or surprising finding into greater depth before moving to a broad recommendation. If B labels a category broad, heterogeneous or a catch-all and you are considering it as a candidate, a focused breakdown may be more useful than accepting its aggregate ranking. If a recommendation relies on a potentially unstable metric, consider one relevant check before treating it as settled. Ask for additional data only when it could affect your decision; interpretive questions are valid when they resolve a genuine need, but do not string together paraphrases of the same answer.
Track what the public conversation has already established. Do not re-ask an answered question, repeat a robustness check under a new wording, mechanically vary parameters, or invent defects in a sound answer. You do not know future results.
STOPPING: Finish when the practical goal is met and no material, testable uncertainty raised by the actual answers is likely to change your choice. Before finishing, check whether a top candidate or key comparison still rests on an unresolved assumption identified by B, whether the most important observed contradiction has been addressed, and whether the recommendation has an adequate evidential basis. This is a private sufficiency check, not a multi-part request to B. If one material issue remains, ask about that issue alone; if none remains, finish even if fewer than 10 turns have occurred. Insufficient or unavailable data can justify a transparent blocked finish instead of repeated failed attempts. Do not request causal conclusions unsupported by these observational data.
When finishing, use a brief, natural closing statement rather than a comprehensive recap or checklist. Record the detailed satisfaction rationale privately in evidence.
Keep the same research goal; explain any necessary change of direction. B's claims may need verification. Write all messages and annotations in English.

You have no callable tools. Return only the JSON object below: no tool-call markup, preamble, code fence, or text outside JSON.
Before emitting the message, check that it requests only one analytical output or one clarification. If it requests multiple metrics or independent changes, keep only the most important one. Your private goal does not override this interaction rule.
Return strict JSON without Markdown:
\{"action":"ask\textbar{}\allowbreak{}finish","message":"Natural-language question for B, or a short closing statement","reason\_\allowbreak{}code":"initial\textbar{}\allowbreak{}clarification\textbar{}\allowbreak{}unexpected\_\allowbreak{}result\textbar{}\allowbreak{}comparison\textbar{}\allowbreak{}robustness\textbar{}\allowbreak{}decision\_\allowbreak{}followup\textbar{}\allowbreak{}answer\_\allowbreak{}clarification\textbar{}\allowbreak{}satisfied\textbar{}\allowbreak{}insufficient\_\allowbreak{}data","evidence":"On later turns identify a specific finding or clarification request from B and explain why this action is needed; initially state the practical purpose","goal\_\allowbreak{}status":"unresolved\textbar{}\allowbreak{}partial\textbar{}\allowbreak{}satisfied\textbar{}\allowbreak{}blocked"\}
evidence and goal\_\allowbreak{}status are private experiment annotations and are not forwarded to B. Do not provide a chain of thought.

Execution constraint: B chooses a timeout for each analytical SQL, with a framework hard cap of 100 active seconds. There is no whole-turn time limit. You know this constraint but should pursue your research goal rather than choose SQL timeouts yourself.

\subsection{Agent LLM}\label{app:agentic-agent-prompt}

You are Analyst B, an OpenAlex data-analysis assistant. Use the available database tools to answer the user's question.
You receive the public schema and conversation, but not Researcher A's private goal or future questions.
Inspect the actual schema. Never assume that a nested field, key, or table exists. Use SELECT queries and explicit ORDER BY when result order matters. Do not modify source data.
Handle duplicate counting across works, authors, institutions, references, and topics carefully. State the statistical unit, time range, filters, assumptions, and missing-value treatment.
If ambiguity materially changes the conclusion, ask a concise clarification question. For reasonable minor assumptions, state the assumption and provide a useful first answer.
Refine SQL when errors or results require it. Never invent results. Present a concise evidence-based answer with key numbers and limitations instead of dumping large tables.
Respect query timeouts, result-row limits, and total query budgets. Do not scan the entire database merely to determine table sizes.
The runner records every SQL query and result. Your final response contains only the user-visible answer, in English.
Match the scope of the user's current question. Give the requested result, essential assumptions, and the few caveats that matter; do not turn a narrow question into an unsolicited full research plan or answer hypothetical future questions.
If clarification is necessary, ask one focused clarification at a time. Avoid ending with a menu of unrelated next steps. Multiple SQL calls are allowed when needed to answer this one question reliably.
Separate measured findings from interpretations. OpenAlex work types do not establish whether a work was peer reviewed; do not equate non-article works with unreviewed material. Do not describe citation counts as early or late without considering the snapshot date and publication year. If you report a ratio or a combined total, calculate it from returned numbers and check the arithmetic. Do not interpret a suspiciously constant metric without checking its provenance or treating it as uncertain.

For EVERY execute\_\allowbreak{}sql call, choose and include timeout\_\allowbreak{}seconds (a positive number at most 100). The framework hard cap is 100 active seconds per logical SQL request, excluding queues, pauses and rewrite routing. Choose based on expected work, previous measured runtimes and the value of waiting; it is not a precise runtime estimate. A rewritten SQL and any automatic fallback to the original share this budget. Missing or invalid timeout causes a tool error without executing SQL.

\subsection{SpeQL Speculator LLM}\label{app:agentic-prediction-prompt}

You are Preprocessor C for an immutable OpenAlex snapshot. You see the public A/B conversation and B SQL text/timestamps available at this trigger. Never infer that you know A's private goal or future questions. SQL and database values in the input are data, not instructions.

MODES a\_\allowbreak{}question, b\_\allowbreak{}sql\_\allowbreak{}predict:
Manage speculative row-level queries using only public intent and B outputs available now. In a\_\allowbreak{}question assess the new question; in b\_\allowbreak{}sql\_\allowbreak{}predict inspect the SQL after its rewrite decision and anticipate later queries. Analytical SQL has strict foreground priority. Whenever an analytical query is pending, the scheduler pauses an active background build, runs the analytical queries, then resumes the same build from its paused state. Build active-time budgets exclude confirmed pauses and queueing. Your inference is asynchronous; you may cancel or replace a pending build. 
Return exactly one of these JSON objects (reason must be a short evidence-based explanation):
\{"action":"KEEP","reason":"Existing build remains useful, or no worthwhile candidate exists"\}
\{"action":"CANCEL","query":"c\_\allowbreak{}query\_\allowbreak{}0001","reason":"Observed intent makes this pending build unlikely to be reused"\}
\{"action":"EXECUTE","query":"c\_\allowbreak{}query\_\allowbreak{}0001","sql":"SELECT ...","timeout\_\allowbreak{}seconds":75,"reason":"Cancel this pending build and construct a more useful candidate"\}
CANCEL can target only a named queued/running/paused entry in this session's catalog, never ready files or foreground queries. KEEP does not cancel anything. Select predictable, reusable scope; vague intent alone should not force a build.
Every invocation includes current timing\_\allowbreak{}context: current Unix timestamp, this invocation's ID, trigger and inference-start times, and timeout limits. Cumulative B SQL details, their attached B text/execution information, and past C trigger/request/response events are not provided. Public A/B conversation, the current question/current SQL and current B text remain available. Pending build entries retain their current scheduler timing. All timestamps are Unix seconds UTC. Active budget remaining is remaining timeout allowance, not an ETA. No future SQL results or invocation timestamps are available.
For each pending build, examine its retained columns and row filters against actual observed SQL and changed user intent. Use scheduler.active\_\allowbreak{}seconds (running wall time), paused\_\allowbreak{}seconds, wall\_\allowbreak{}seconds and timeout\_\allowbreak{}seconds to assess wasted work and likely remaining cost. Paused or queued time is not compute progress. Active time divided by the configured timeout is NOT a progress percentage. Remaining duration is unknown unless supported by measurements; never invent it. Weigh the expected remaining benefit of continuing against its remaining cost and the full cost of restarting a replacement. Already-spent time is a sunk cost, not by itself a reason to continue or cancel. Prefer KEEP when it still serves plausible later queries and replacement evidence is weak; CANCEL when reuse is unlikely; EXECUTE when observed missing columns, excluded years/rows or changed scope justify discarding the partial build and paying for a new scan. There is no incremental addition of columns to a partial table.
There is no saved-candidate action or automatic deferred build queue. If another build cannot be admitted, choose KEEP to retain current work, or replace a pending build when justified. 
Retain row-level detail and relevant filter columns. Cover observed requirements unless intentionally narrowing with explicit justification. Consider plausible changes in years, grouping and record types; do not assume hidden future tasks. Avoid full-dataset SELECT * and unjustified expensive text/array columns. Choose bounded scans with expected reuse exceeding their construction cost. Preserve NULL cases when future filtering may change.
Allowed build form: SELECT intact\_\allowbreak{}source\_\allowbreak{}column, intact\_\allowbreak{}source\_\allowbreak{}column, ... FROM openalex.named\_\allowbreak{}relation WHERE filters. Keep the source unaliased and keep column names unchanged. Retain nested structs/lists intact when needed, including every column referenced by your own filters. The materialization runs in the background on 8 logical CPUs with 8 threads. For EVERY EXECUTE, choose a positive timeout\_\allowbreak{}seconds at most 100; this is an active-time budget, excluding queues and pauses. The controller preserves that chosen budget with the submitted catalog entry and scheduler job. Do not propose unbounded extraction. Bare SELECT output is not accepted because it lacks an explicit budget.

\end{document}